\documentclass[twocolumn]{jpsj2} 
\usepackage{bm}
\usepackage{tabularx}
\title{Complex-Orbital Order in Fe$_3$O$_4$ and Mechanism of the Verwey Transition}
\author{
Hisashi \textsc{Uzu} and Arata \textsc{Tanaka}\thanks{E-mail address: atanaka@hiroshima-u.ac.jp}
}
\inst{
Department of Quantum Matters, ADSM, Hiroshima University, Higashi-Hiroshima 739-8530, Japan
}
\abst{

Electronic state and the Verwey transition in magnetite (Fe$_3$O$_4$) are studied using a spinless three-band Hubbard model for 3$d$ electrons on the $B$ sites with the Hartree-Fock approximation and the exact diagonalisation method. 
Complex-orbital, e.g., $\frac{1}{\sqrt{2}}[|zx\rangle+\textrm{i}|yz\rangle]$, ordered (COO) states having noncollinear orbital moments $\sim 0.4~\mu_{\rm B}$ on the $B$ sites are obtained with the cubic lattice structure of the high-temperature phase.  The COO state is a novel form of magnetic ordering within the orbital degree of freedom. It arises from the formation of Hund's second rule states of spinless pseudo-$d$  molecular orbitals in the Fe$_4$ tetrahedral units of the $B$ sites and ferromagnetic alignment of their fictitious orbital moments.
A COO state with longer periodicity is obtained with pseudo-orthorhombic {\it Pmca} and {\it Pmc}\,2$_1$ structures for the low-temperature phase. The state spontaneously lowers the crystal symmetry to the monoclinic and explains experimentally observed rhombohedral cell deformation and Jahn-Teller like distortion. From these findings, we consider that at the Verwey transition temperature, the COO state remaining to be short-range order impeded by dynamical lattice distortion in high temperature is developed into that with long-range order coupled with the monoclinic lattice distortion.
}
\kword{magnetite, Fe$_{3}$O$_{4}$, the Verwey transition, complex-orbital order, charge order, noncollinear orbital moment, pyrochlore lattice, multiferroic}
\begin{document}
\maketitle
\section{Introduction}
Magnetite (Fe$_{3}$O$_{4}$) is one of the most familiar minerals for human beings. Known as a loadstone even in the Greek era, it has been attracting us for its magnetic properties. In 1939, Verwey discovered an abrupt increase of resistivity in magnetite by two orders of magnitude on cooling it below $T_{\rm V} \sim 120$~K~\cite{discovery}.  Despite numerous experimental and theoretical studies, the mechanism of this Verwey transition has been controversial for more than half a century.

\begin{figure}[tb]
\begin{center}
\includegraphics[width=7.5cm]{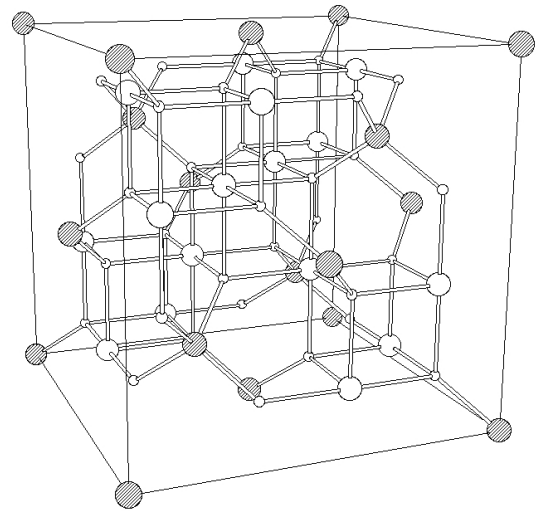}
\end{center}
\caption{Illustration of crystal structure of Fe$_3$O$_4$ in the high-temperature cubic cell. The large open and hatched circles represent Fe ions on the $B$ and $A$ sites, respectively and small circles denote oxygen ions.
} 
\label{Fe3O4}
\end{figure}

Magnetite is a kind of spinel ferrites. It crystallises in the cubic inverse spinel structure at room temperature (see Fig.~\ref{Fe3O4}). One-third of Fe ions are on tetrahedral {\it A} sites with a Fe$^{3+}$ formal valence constituting a diamond lattice. The remaining two-thirds of Fe ions are on octahedral {\it B} sites in a Fe$^{2.5+}$ formal average valence state forming a pyrochlore lattice. Below $T_{\rm N}\sim 860$~K magnetite is a ferrimagnet, where the magnetic moments of Fe ions on the $A$ and $B$ sites are antiparallel and the net magnetisation is $\sim 4~\mu_{\rm B}$ per unit formula.
The tetrahedral (octahedral) oxygen coordination of the $A$ ($B$) site Fe ion causes ligand field splitting of the 3$d$ level into three-fold $t_{2g}$ and two-fold $e_g$ levels. 
The 3$d$ orbitals with minority (majority) spin on the $A$ ($B$) site are fully occupied and half of Fe ions on the $B$ sites accommodate an extra electron with minority spin in their $t_{2g}$ orbital in each.

The Verwey transition is a first-order phase transition accompanied by a lattice deformation from the cubic ({\it Fd}\=3{\it m}) symmetry. Below $T_\textrm{V}$, the cubic cell is elongated to nearly rhombohedral along the [111]$_{\rm c}$ direction (the subscript `c' denotes the cubic cell). The lattice symmetry of the low-temperature phase is still unresolved problem.
X-ray and neutron diffraction studies~\cite{MIizumimono,WrightLett,WrightPRB} have shown that the low-temperature structure has a monoclinic {\it Cc} symmetry with a $\sqrt{2}a_{\rm c}\times \sqrt{2}a_{\rm c}\times 2a_{\rm c}$ crystallographic supercell.  Small triclinic distortion from the $Cc$ structure has been inferred from the observation of the magnetoelectric effect~\cite{YMiyamoto1,YMiyamoto2} and a study of twins with the synchrotron radiation x-ray topography~\cite{CMedrano}. However, no evidence has been found for the breaking of the inversion symmetry in an infrared spectroscopic investigation~\cite{LVGasparov}.

 Verwey and co-workers have interpreted the transition as an order-to-disorder transition of the Fe$^{2+}$ and Fe$^{3+}$ ions on the $B$ sites; Fe$^{2+}$ and Fe$^{3+}$ layers stack alternately along the [001]$_{\rm c}$ direction so as to minimise Coulomb energy in their charge-ordering model~\cite{VerweyCO}. 
Although intensive experimental and theoretical efforts have been made, the existence of the charge ordering in the low-temperature phase as well as its electronic states above and below $T_\textrm{V}$ still remain matters of debate~\cite{NTsuda,JGarcia,FWalz}.

Several theoretical models  have been proposed to explain  the mechanism of the Verwey transition.
Cullen and Callen have discussed the transition using a spinless one-band Hubbard model within the Hartree approximation~\cite{CullenCallen}. They assumed that the trigonal crystal-field is strong enough to lift all the degeneracy of the $t_{2g}$ orbitals and only the low-lying $a_{1g}$ orbital of every Fe ion on the $B$ sites is taken into account. They found that a charge-ordered state is stabilised when Coulomb interaction between the neighbouring sites exceeds a critical value. Mishra \textit{et al.} have extended the one-band model to a three-band model considering the triple degeneracy of the $t_{2g}$ orbitals~\cite{SKMishra}. Their results are similar to those of Cullen and Callen's except for the ferro-orbital ordering, where only one kind of the $t_{2g}$ orbitals ($xy$, $yz$ or $zx$ orbitals) is taken for all the occupied sites~\cite{HSeo}. Based on the three-band model, Seo \textit{et al.} have proposed a mechanism of the Verwey transition with a bond dimerisation caused by the Peierls instability, where charge ordering is irrelevant to the transition~\cite{HSeo}. 
Recently, the electronic state in the low-temperature phase has been investigated using local density approximation + Hubbard U (LDA + {\it U}) calculations~\cite{ILeonovLett, HTJengLett, ILeonovFull, HTJengFull} based on the monoclinic lattice structure shown by Wright \textit{et al}~\cite{WrightPRB}. In the LDA + {\it U} calculations, a state with simultaneous charge and antiferro-orbital ordering is obtained. 

  Large orbital moments at the $B$ sites above and below $T_{\rm V}$ have been observed in recent measurements of Fe $L_{2,3}$-edge x-ray magnetic circular dichroism (XMCD)~\cite{DJHuang}.
  However, more recent XMCD experiments have resulted in observing no net orbital moments~\cite{Vanishing}. On the other hand, orbital moment $\sim 0.5~\mu_{\rm B}$ has been also inferred from the magnetic Compton scattering experiments~\cite{Yinwan}. 

In previous works, we have investigated a spinless three-band Hubbard model with the Hartree-Fock (HF) approximation and exact diagonalisation method and found complex-orbital ordered (COO) states with noncollinear orbital moments as the ground states within a realistic parameter range~\cite{PhysicaB,JPSJLett}. Here, the word `complex'-orbital represents an orbital which is described as a linear combinations of the $t_{2g}$ basis functions ($xy$, $yz$ and $zx$ orbitals) with \textit{complex number} coefficients and has non-zero orbital moment.

The purpose of this paper is to describe the formation mechanism of the COO state and a possible COO state realised in the low-temperature phase and to discuss the mechanism of the Verwey transition. We found that the COO state is a kind of magnetic state within the orbital degree of freedom. It does not originate from the mechanisms in the conventional orbital order, which can be described with the Kugel-Khomskii type model~\cite{KIKugel} or in usual magnetic materials, where orbital moments are induced by ordered spin via the spin-orbit interaction. 
A COO state with longer periodicity is expected to be stabilised in the low-temperature phase owning to the strong coupling between the orbital polarisation and lattice distortion. The state  
explains experimentally observed rhombohedral cell deformation and Jahn-Teller like distortion.

This paper is organised as follows. 
The details of the model Hamiltonian are presented in \S~\ref{Model}.
Charge and orbital ordered states found in the HF calculations with the cubic structure ($T>T_{\rm V}$) are described in \S~\ref{HighT}. The formation mechanism of the COO state is discussed analytically and the stability of the state against the electron correlation effects is examined with the exact diagonalisation method for a 16-site model in \S~\ref{Mechanism}. After short introduction on the lattice structure and related properties in $T<T_{\rm V}$, discussion on the COO state in connection with lattice distortion and description on a  COO state obtained from the HF calculations with pseudo-orthorhombic structure models for $T<T_{\rm V}$ are provided in \S~\ref{LowT}. In \S~\ref{Discussions}, electronic states above and below $T_{\rm V}$ and the mechanism of the Verwey transition are discussed. Finally, a summary is given in \S~\ref{Conclusions}.
%
\section{Three-Band Hubbard Model\label{Model}}
To discuss possible charge and orbital orderings in magnetite, we studied  a spinless three-band Hubbard model for 3$d$ electrons on the pyrochlore lattice of the $B$ sites. The pyrochlore lattice is a network of corner-shared tetrahedra consisting of  linear chains with six different directions $[1,\pm 1,0]_{\rm c}$, $[1,0,\pm 1]_{\rm c}$ and  $[0,1,\pm 1]_{\rm c}$ (see Fig.~\ref{Fe3O4}). This model was originally introduced by Mishra \textit{et al.}~\cite{SKMishra} as an extension of the one-band Hubbard model of Cullen and Callen~\cite{CullenCallen}. In the three-band model, only $t_{2g}$ electrons with minority spins at the {\it B} sites are considered on the assumption that the spin moments at the {\it B} sites are collinear, and the 3$d$ orbitals with majority spins are fully occupied. We furthermore introduce effects of the spin-orbit interaction, a trigonal crystal-field  and electron hopping via adjacent O 2$p$ orbitals. As shown later, electron hopping among the {\it B} sites via the neighbouring oxygen ions is indispensable for the complex-orbital orderings. 

The Hamiltonian for $t_{2g}$ electrons with minority spins on the {\it B} sites is written as
\begin{align}
H&=\sum_{\langle ij\rangle,\mu\nu}t_{ij}^{\mu\nu}(c_{i\mu}^\dagger c_{j\nu}+c_{j\nu}^\dagger c_{i\mu}) \nonumber\\
& +\sum_{i,\mu>\nu}Un_{i\mu}n_{i\nu}+\sum_{\langle ij\rangle,\mu\nu}Vn_{i\mu}n_{j\nu} \nonumber\\
& + \sum_{i,\mu\nu}c_{i\mu}^\dagger\Bigl[D\{(\bm{\alpha}_i \cdot \bm{l} )^2 -\frac{2}{3}\}- \frac{\zeta}{2}(\bm{\alpha}_{\rm S}\cdot\bm{l})\Bigr]c_{i\nu}. \label{eq1}
\end{align}
Here $t_{ij}^{\mu\nu}$ denotes hopping integral between orbital $\mu$ (=$xy$, $yz$ or $zx$) at site $i$ and orbital $\nu$ at site $j$. $c_{i\mu}^\dagger$, $c_{i\mu}$ and $n_{i\mu}$ represent the creation, annihilation and number operators of an electron on orbital $\mu$ at site $i$, respectively. $\langle ij\rangle$ stands for summation over the nearest-neighbour sites. $U$ and $V$ are the on-site and inter-site Coulomb energies, respectively. 
The first term in the square brackets on the third line describes the trigonal crystal field, where $\bm{l}$ is an $l=1$ pseudo-angular momentum operator, and the unit vector $\bm{\alpha}_i$ is along the three-fold axis of site $i$. The trigonal field splits the $t_{2g}$ levels at a $B$ site into a single $a_{1g}$ and two-fold $e^\pi_{g}$ levels, and the level splitting  is denoted as $D$($=\varepsilon(e^\pi_{g})-\varepsilon(a_{1g})$).  
The second term in the square brackets is the spin-orbit interaction with a coupling constant $\zeta=0.052$~eV; the unit vector $\bm{\alpha}_{\rm S}$ represents the direction of the minority spins at the {\it B} sites. 

Hopping integrals between nearest-neighbour sites $i$ and $j$ can be describe with the Slater-Koster parameters $t_{dd\sigma}$, $t_{dd\pi}$ and $t_{dd\delta}$ as~\cite{WAHarrison}
\begin{equation}
\begin{split}
t_{ij}^{xy,xy}&=3l^2m^2t_{dd\sigma}+(l^2+m^2-4l^2m^2)t_{dd\pi},\\
t_{ij}^{yz,zx}&=3lmn^2t_{dd\sigma}+lm(1-4n^2)t_{dd\pi}- {\rm sgn}(lm)\,t_{pd},\\
\end{split}
\end{equation}
where $l$, $m$ and $n$ denote direction cosines ($l$, $m$, $n$) along site $i$ to site $j$ direction and terms with $t_{dd\delta}$ are omitted.  The additional term with $t_{pd}$ in $t_{ij}^{yz,zx}$ describes electron hopping through the adjacent O 2$p_z$ orbital.  There is the relation $t_{ij}^{\mu,\nu}=t_{ij}^{\nu,\mu}$, and one can obtain all other terms by interchanging the $x$, $y$ and $z$ coordinates.
The relations $t_{dd\sigma}=-63.2r^{-5}$~eV and $t_{dd\pi}=34.1r^{-5}$~eV are assumed after Harrison\cite{WAHarrison}, where $r$ (in units of \AA) represents the distance between sites $i$ and $j$.

The values of hopping integrals for the $B$ sites along the $[1,\pm1,0]_{\rm c}$ chains in the cubic structure are $t_{ij}^{xy,xy}=3t_{dd\sigma}/4=-0.206$~eV, $t_{ij}^{yz,yz}=t_{ij}^{zx,zx}=t_{dd\pi}/2=0.074$~eV and $t_{ij}^{yz,zx}$=$t_{ij}^{zx,yz}=\pm(t_{dd\pi}/2- t_{pd}) =\pm (0.074~{\rm eV} - t_{pd})$, and all other terms are zero. 
$t_{pd}$ corresponds to the hopping process: $\textrm{Fe}^{3+}(3d^5);\textrm{O}^{2-}(2p^6);\textrm{Fe}^{2+}(3d^6)$ $\to$ \\ $\textrm{Fe}^{2+}(3d^6);\textrm{O}^{-}(2p^5);\textrm{Fe}^{2+}(3d^6)$ $\to$ $\textrm{Fe}^{2+}(3d^6);\textrm{O}^{2-}(2p^6);\textrm{Fe}^{3+}(3d^5)$ and the value is roughly estimated to be $t_{pd}= t_{pd\pi}^2/\Delta \sim 0.36$~eV by means of the second-order perturbation theory. Here, $t_{pd\pi}\sim 0.6$~eV is the Slater-Koster parameter between Fe 3$d$ and O 2$p$ orbitals, and $\Delta\sim 1$~eV is the charge-transfer energy at the Fe$^{3+}$ ion site defined  as the averaged energy difference of the 3$d^6$ with a hole on the ligand and the 3$d^5$ configurations~\cite{JChen}.
The value of $t_{pd}$ is comparable to the direct $d$-$d$ hopping terms and not negligible. 
$U=4.0$~eV is adopted~\cite{ZZhang}. We do not know the accurate value of $D$, and here, $D=0.25$~eV is assumed; the $D$ dependence of electronic state will be discussed in \S~4.2.

\begin{figure*}[htb]
\begin{center}
\includegraphics[width=7cm]{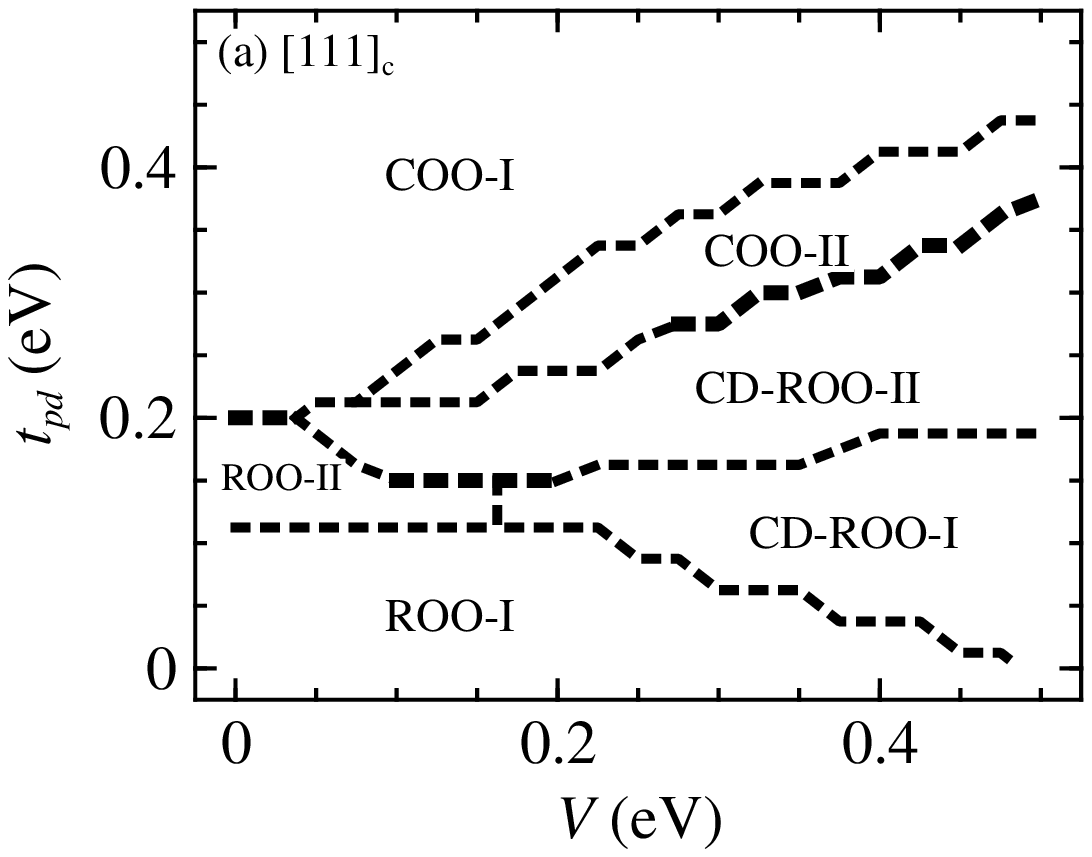}~~~\includegraphics[width=7cm]{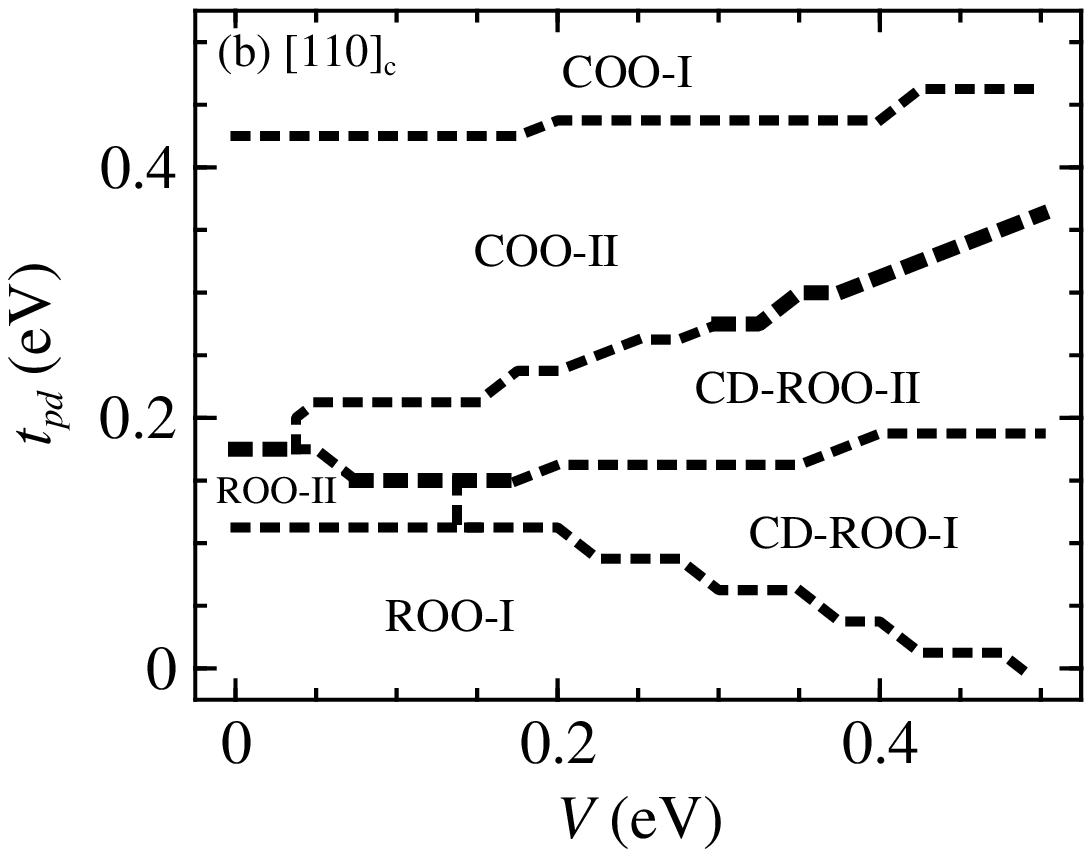}\\[3mm]
\includegraphics[width=7cm]{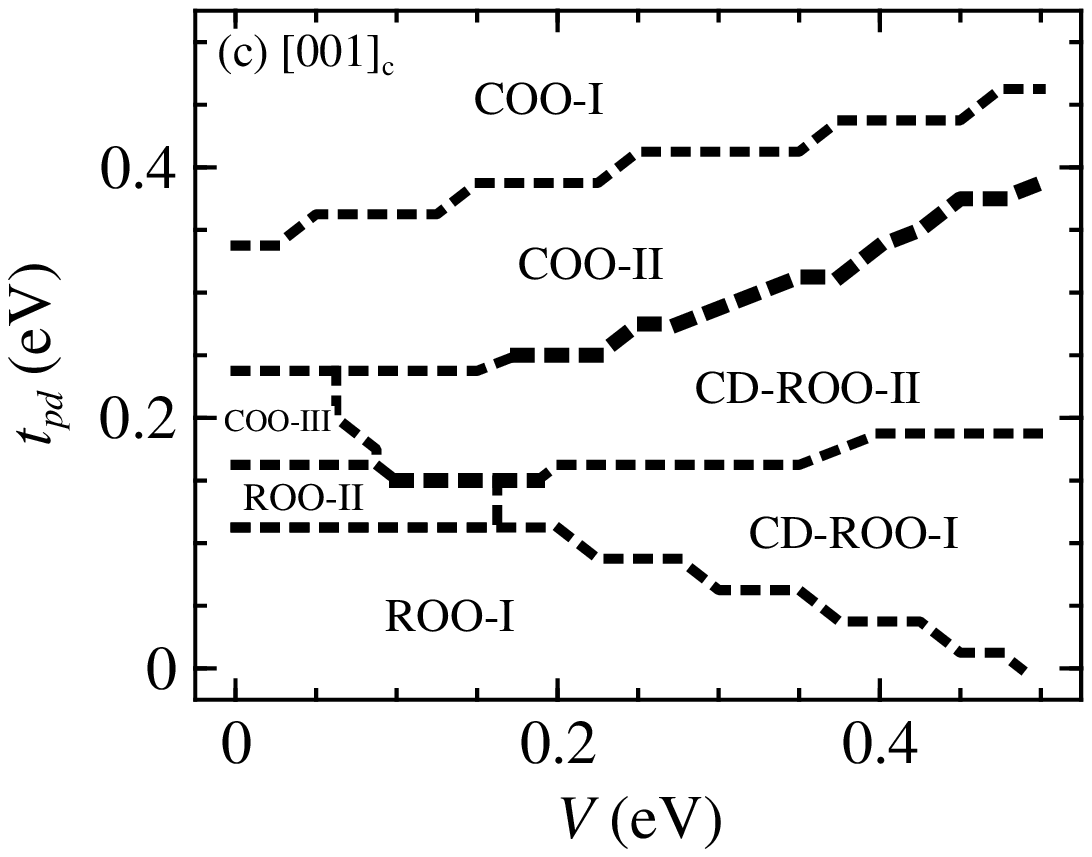}~~~\includegraphics[width=7cm]{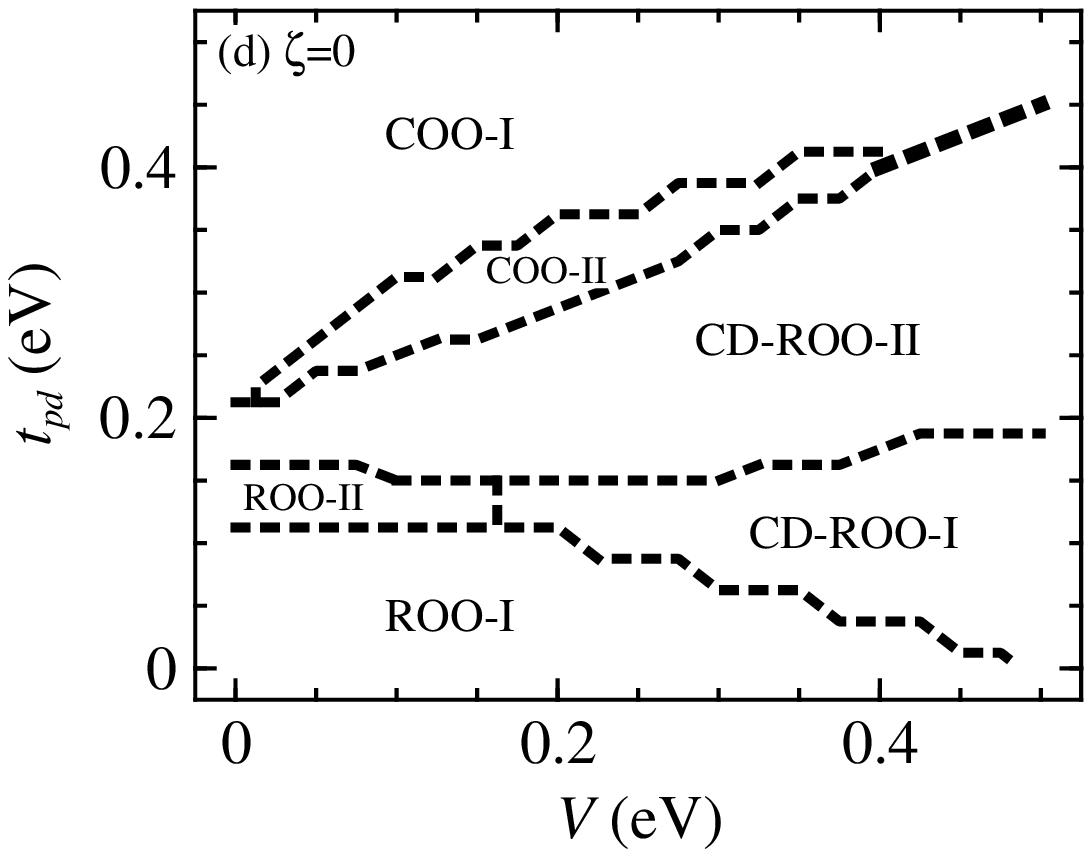}
\end{center}
\caption{
Phase diagrams of the ground states for the high-temperature phase in $V$-$t_{pd}$  plane with $D=0.25$~eV. In the calculations, the directions of minority spins are assumed along directions [111]$_{\rm c}$ (a), [110]$_{\rm c}$ (b) and [001]$_{\rm c}$ (c). The same but without the spin-orbit interaction is also shown in (d) for comparison.
`ROO' and `COO' represent real- and complex- orbital ordered states, respectively, and `CD' denotes the state with the Verwey charge ordering. Weak charge disproportionation occurs in the COO-I and COO-II states (for details see text), and the charge distributions in the ROO-I, ROO-II and COO-III states are uniform.
} 
\label{PD1}
\end{figure*}
\section{Electronic State in the Cubic Structure\label{HighT}}
Electronic structure calculations for the high-temperature phase were performed within the HF approximation assuming the cubic fcc unit cell with four independent $B$ sites (see Fig.~\ref{Fe3O4}).  The wave vectors throughout the first Brillouin zone of the fcc lattice were sampled using the tetrahedron method~\cite{TM}. 
 In the HF approximation, we have
\begin{align}
 n_{i\mu} n_{j\nu} \rightarrow & n_{i\mu} \langle n_{j\nu} \rangle + \langle n_{i\mu} \rangle n_{j\nu}- \langle n_{i\mu} \rangle \langle n_{j\nu} \rangle  \nonumber\\
& - c^\dagger_{i\mu}c_{j\nu}\langle c^\dagger_{j\nu}c_{i\mu} \rangle- \langle c^\dagger_{i\mu}c_{j\nu} \rangle c^\dagger_{j\nu}c_{i\mu} \nonumber\\
& +\langle c^\dagger_{i\mu}c_{j\nu} \rangle \langle c^\dagger_{j\nu}c_{i\mu} \rangle. \label{HF}
\end{align}
Only the Hartree terms [the first three terms in  eq.~(\ref{HF})] were considered in the previous studies\cite{SKMishra,HSeo}; however, inclusion of the Fock terms [the last three terms in eq.~(\ref{HF})] are essential to keep the spherical symmetry of the Coulomb interaction. While orbital orderings with variety of orbitals described by linear combinations of the three $t_{2g}$ orbitals are obtained with the Fock terms, limited orderings with pure $yz$, $zx$ or $xy$ orbitals are found without the term.

In Fig.~\ref{PD1}, phase diagrams obtained with $D=0.25$~eV assuming the directions of minority spins along directions [111]$_{\rm c}$ (a), [110]$_{\rm c}$ (b) and [001]$_{\rm c}$ (c) are shown.
For comparison, that calculated without the spin-orbit interaction ($\zeta=0$) is also depicted in (d). 
In each panels,  six different phases COO-I, COO-II, ROO-I, ROO-II, CD-ROO-I and CD-ROO-II can be seen, except for the additional small area of COO-III phase in Fig.~1(c). These phases can be classified into two groups whether ordered orbitals are described as linear combinations of the $xy$, $yz$ and $zx$ orbitals with {\it real number} coefficients (labelled with `ROO') or {\it complex number} coefficients (labelled with `COO'). 
The CD-ROO-I and CD-ROO-II states are real-number orbital-ordered (ROO) states accompanied by charge ordering where the layers of Fe$^{2+}$ and Fe$^{3+}$ ions are alternately piled up along the [001]$_c$ direction. For small values of $t_{pd}$ the ROO states are stable, whereas the complex-number orbital-ordered (COO) states are favourable for large $t_{pd}$ values. The ROO states with charge ordering prevail with large values of $V$. Differences among the phase diagrams are rather limited; large variations can be found only in the boundary between the COO-I and COO-II phases and the appearance of the COO-III phase with $\bm{\alpha}_{\rm S}//[001]_c$.

The $a_{1g}$ orbital of every $B$ site is occupied by 0.5 electron in the ROO-I state, a similar state but with an additional ferro-orbital component (one of the $xy$, $yz$ and $zx$ orbitals) is stabilised in the ROO-II state. 
For $D=0$~eV, we found the ferro-orbital ordered state corresponding to that discussed in the Hartree calculations in ref.~\citen{HSeo} with small value of $V$, in the range of 0.05~eV $\le t_{pd}\le 0.15$~eV. The ROO-II emerges as a result of mixture of the ferro-orbital ordering and the $a_{1g}$ orbital ordered state that are strengthen by the trigonal fields ($D=0.25$~eV).  The CD-ROO-I and CD-ROO-II states are accompanied by the Verwey charge orderings, where the layers of Fe$^{2+}$ and Fe$^{3+}$ ions are alternately piled up along the [001]$_c$ direction. The $a_{1g}$ orbital of every Fe$^{2+}$ ion site is occupied in the CD-ROO-I state. In the CD-ROO-II state, $\frac{1}{\sqrt{2}}[|yz\rangle+|zx\rangle]$ orbitals are occupied in Fe$^{2+}$ ions on [110]$_{\rm c}$ chains as the lobes of the orbitals are parallel to the Fe$^{2+}$ chains. The ROO-I and CD-ROO-I states found in the small $t_{pd}$ regime are essentially the same as those discussed by Cullen and Callen\cite{CullenCallen}, where the orbital degeneracy is lifted. 

\begin{figure*}[t]
\begin{center}
\includegraphics[width=3.8cm]{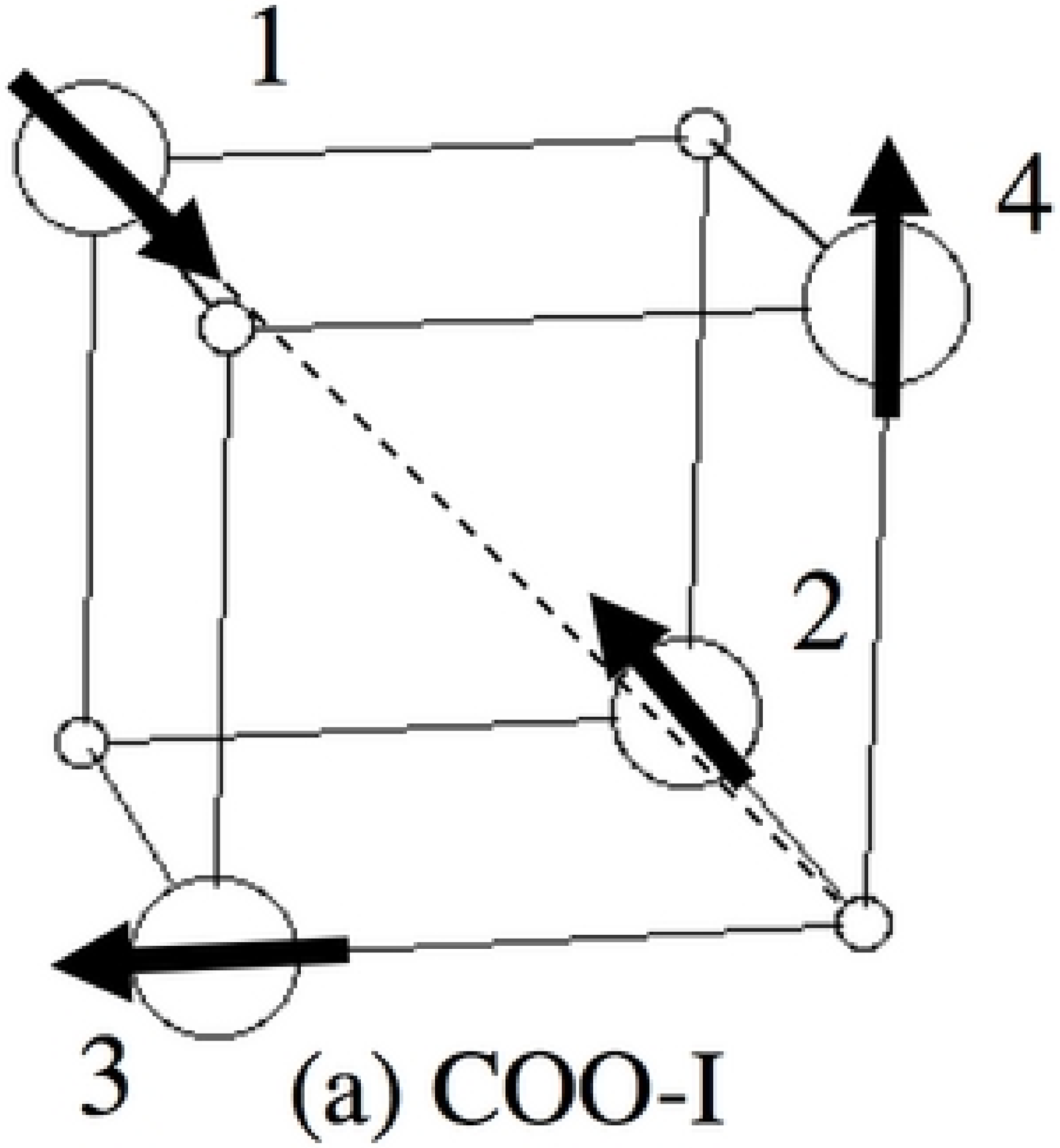}~~~\includegraphics[width=3.8cm]{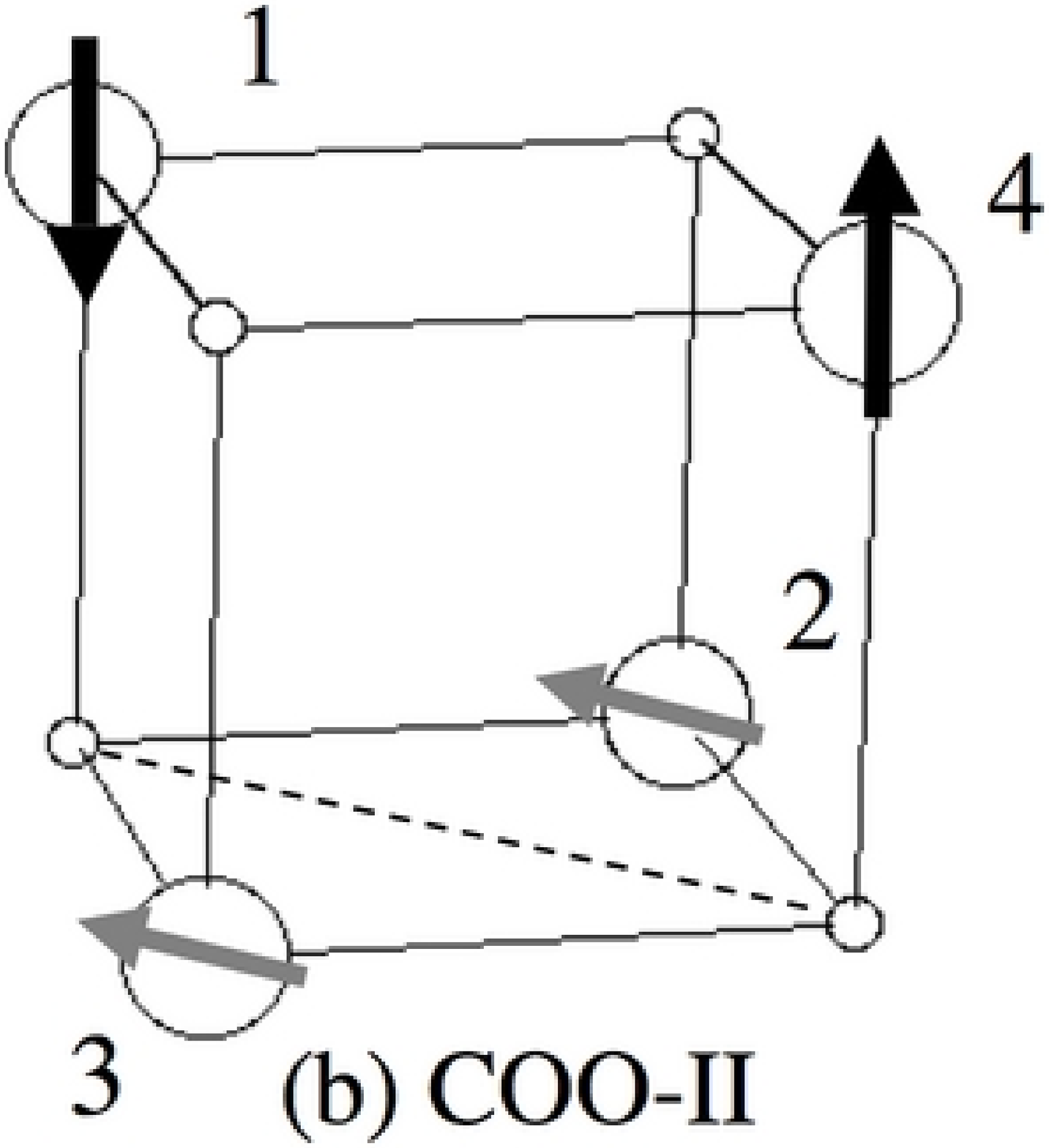}~~~\includegraphics[width=3.8cm]{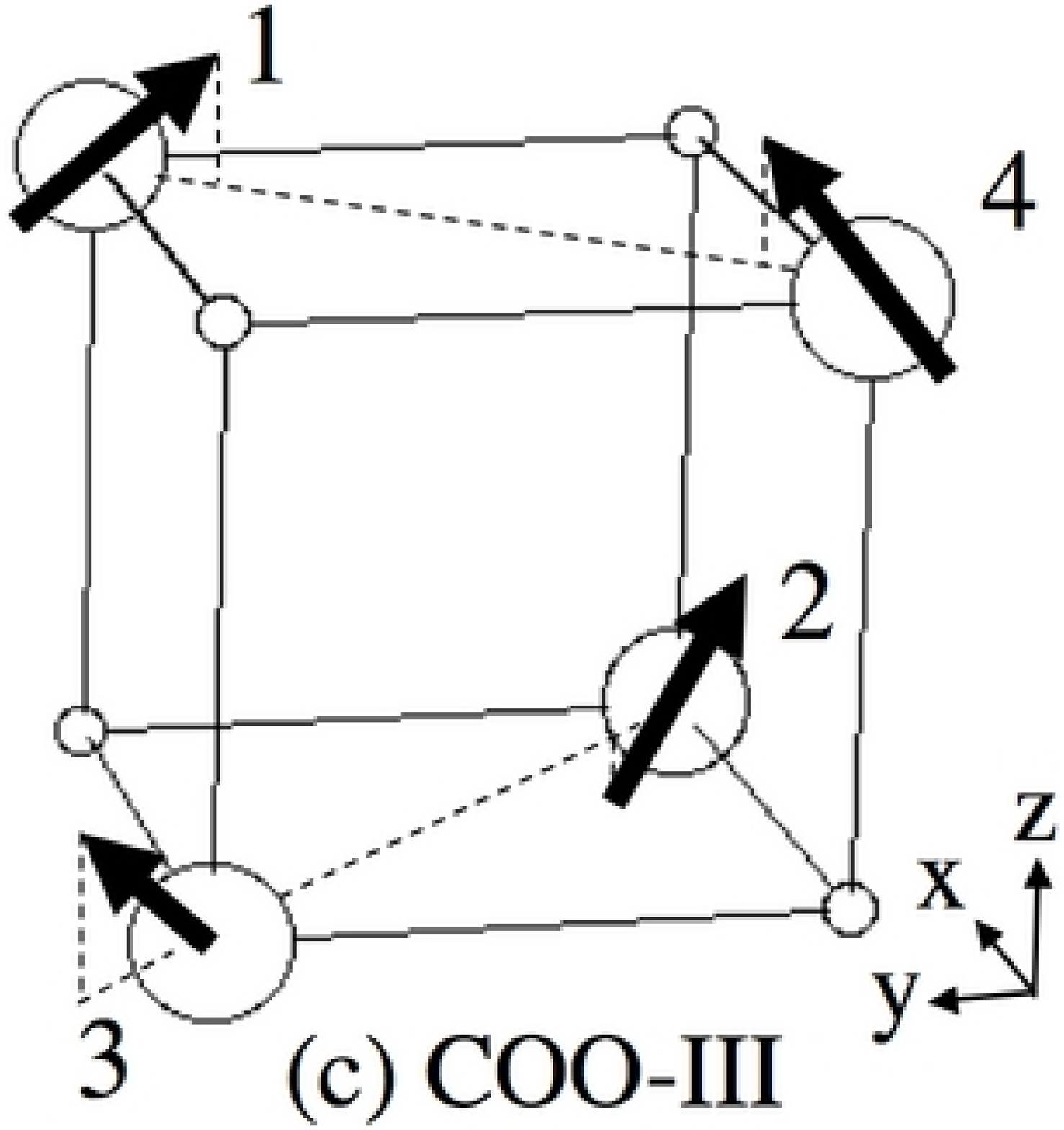}
\end{center}
\caption{
Alignment of orbital moments in Fe$_4$O$_4$ cube for the COO-I (a), COO-II (b) and COO-III (c) states. The large and small circles signify Fe ions at the $B$ sites and oxygen ions, respectively. The arrows on the four independent $B$ sites labelled 1 - 4 denote the directions of the orbital moments. The grey (black) arrows represent electron-poor (electron-rich) sites in (b). The magnetic easy axes of the COO-I, COO-II and COO-III states are oriented along the [111]$_{\rm c}$, [110]$_{\rm c}$ and [001]$_{\rm c}$ directions, respectively. Details of the electronic states on each of the COO states are listed in Table~\ref{tbcoo}. 
} 
\label{coo}
\end{figure*}
\begin{table*}[thb]
\caption{
Details of the COO-I, COO-II and COO-III states obtained with the high-temperature cubic phase. The direction of trigonal axis $\bm{\alpha}_i$, dominantly occupied orbital (with $\sim 0.5$ electron) and its orbital moment (in units of $\mu_{\rm B}$) on each of four independent $B$ sites (see Fig.~\ref{coo}) are listed for each COO state.
}
\vspace{2mm}
\begin{center}
\extrarowheight=0.3mm
\begin{tabular}{ccclc}\hline
COO state & site & $\bm{\alpha}_i$ &~~~~orbital [ $\omega=\exp({2\pi \textrm{i}/3})$ ] & moment   \\ \hline
& 1 & [111]$_{\rm c}$ & $|e^{-}\rangle~=\frac{-1}{\sqrt{3}}[|xy\rangle+\omega|yz\rangle+\omega^2|zx\rangle]$ & $\frac{-1}{\sqrt{3}}(1,1,1)$  \\ 
\raisebox{-2mm}[0pt][0pt]{COO-I} & 2 & [1\=1\=1]$_{\rm c}$ & $|x^{+}\rangle~= \frac{1}{\sqrt{2}}[|zx\rangle-\textrm{i}|xy\rangle]$ & $(1,0,0)$ \\ 
 & 3 & [\=11\=1]$_{\rm c}$ & $|y^{+}\rangle~=\frac{1}{\sqrt{2}}[|xy\rangle-\textrm{i}|yz\rangle]$ & $(0,1,0)$ \\  
& 4 & [\=1\=11]$_{\rm c}$ & $|z^{+}\rangle~=\frac{1}{\sqrt{2}}[|yz\rangle-\textrm{i}|zx\rangle]$ & $(0,0,1)$ \\ \hline
& 1 & [111]$_{\rm c}$ & $|z^{-}\rangle~=\frac{1}{\sqrt{2}}[|yz\rangle+\textrm{i}|zx\rangle]$ & $(0,0,-1)$ \\ 
\raisebox{-2mm}[0pt][0pt]{COO-II} & 2 & [1\=1\=1]$_{\rm c}$ & $|\textrm{110}\rangle=\frac{1}{\sqrt{6}}[2|xy\rangle-\textrm{i}|yz\rangle+\textrm{i}|zx\rangle]$ & $\frac{2}{3}(1,1,0)$ \\ 
& 3 & [\=11\=1]$_{\rm c}$ & $|\textrm{110}\rangle=\frac{1}{\sqrt{6}}[2|xy\rangle-\textrm{i}|yz\rangle+\textrm{i}|zx\rangle]$& $\frac{2}{3}(1,1,0)$ \\  
& 4 & [\=1\=11]$_{\rm c}$ & $|z^{+}\rangle~=\frac{1}{\sqrt{2}}[|yz\rangle-\textrm{i}|zx\rangle]$ & $(0,0,1)$ \\ \hline
& 1 & [111]$_{\rm c}$ & $|\textrm{\=1\=11}\rangle=\frac{1}{\sqrt{3}}[|xy\rangle-\omega^2|yz\rangle-\omega|zx\rangle]$ & $\frac{-1}{\sqrt{3}}(1,1,-1)$ \\ 
\raisebox{-2mm}[0pt][0pt]{COO-III} & 2 & [1\=1\=1]$_{\rm c}$ & $|\textrm{1\=11}\rangle=\frac{1}{\sqrt{3}}[|xy\rangle+\omega|yz\rangle-\omega^2|zx\rangle]$ & $\frac{1}{\sqrt{3}}(1,-1,1)$ \\ 
& 3 & [\=11\=1]$_{\rm c}$ & $|\textrm{\=111}\rangle=\frac{1}{\sqrt{3}}[|xy\rangle-\omega|yz\rangle+\omega^2|zx\rangle]$ & $\frac{1}{\sqrt{3}}(-1,1,1)$ \\  
& 4 & [\=1\=11]$_{\rm c}$ & $|\textrm{111}\rangle=\frac{1}{\sqrt{3}}[|xy\rangle+\omega^2|yz\rangle+\omega|zx\rangle]$ & $\frac{1}{\sqrt{3}}(1,1,1)$ \\ \hline
\end{tabular}
\end{center}
\label{tbcoo} 
\end{table*}
Figure~\ref{coo} shows the directions of the orbital moments in the COO-I, COO-II and COO-III states.  The chiefly occupied $t_{2g}$ orbitals  ($\sim 0.5$ electron per site) and their orbital moments are listed in Table~\ref{tbcoo}. 
The total orbital moments are along the [111]$_c$, [110]$_c$ and [001]$_c$ directions  in the COO-I, COO-II and COO-III states, respectively. These COO states have the same periodicity to the crystal, but the orbital ordering spontaneously lower the space group symmetry from cubic ($Fd\overline{3}m$) to rhombohedral ($R\underline{\overline{3}}\underline{m}$), orthorhombic ($Im\underline{m}\underline{2}$) and tetragonal ($I\underline{\overline{4}}\underline{2}\underline{m}$), respectively. Because of noncollinear alignment of the moments, the size of total orbital moments are rather small. For instance, in the COO-I state, each $B$ site has orbital moment of about $0.5~\mu_\textrm{B}$. However, the total orbital moment is only $\sim 0.20~\mu_\textrm{B}$ per $B$ site, since the orbital moment at site 1 is in the opposite direction to the sum of those at sites 2, 3 and 4, and this almost cancel out the net orbital moment of each Fe$_4$O$_4$ cube. As was mentioned before, the spin-orbit interaction is not essential ingredient for the stability of the COO states. This becomes evident when the results obtained with [see Figs.~\ref{PD1}(a)-(c)] and without [see Fig.~\ref{PD1}(d)] the spin-orbit interaction are compared, where the phase boundaries between the COO and ROO states are almost unchanged among the diagrams. On the other hand, relative stability among the three COO states are affected by the direction of the spin moment, since the differences in the total energies among the three are only $\sim$0.01~eV per site. Indeed, as clearly seen in Figs.~\ref{PD1}(a)-(d), each of the COO states is more stabilised when $\bm{\alpha}_{\rm S}$ is along the direction of its total orbital moment as compared the other directions. 

The charge disproportionation of the COO states is rather moderate as compared to the CD-ROO-I or CD-ROO-II states.
The charge disproportionation of the COO-II state approaches uniform with decreasing value of $V$ and increasing $t_{pd}$, and the differences in the occupation number between electron-poor and electron--rich sites are  from 0.05 to 0.29 in the phase diagram of Fig.~\ref{PD1}(b). On the other hand, the COO-I state is considerably uniform, and the differences in the occupation number among sites are at most 0.09 in the phase diagram of Fig.~\ref{PD1}(a).

\begin{figure}[tb]
\begin{center}
\includegraphics[width=8.4cm]{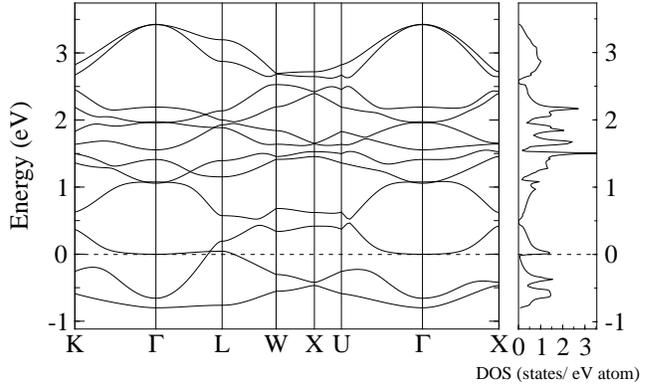}
\end{center}
\caption{
Band structure and the density of states for the COO-I state obtained with $V=0.1$~eV, $t_{pd}=0.35$~eV, $D=0.25$~eV and $\bm{\alpha}_{\rm S}=\frac{-1}{\sqrt{3}}(1,1,1)$. The broken lines in the figure denote the Fermi energy. The symmetry points K$(\frac{3}{4},0,\frac{3}{4})$, L$(\frac{1}{2},\frac{1}{2},\frac{1}{2})$, U$(\frac{1}{4},\frac{1}{4},1)$, W$(\frac{1}{2},0,1)$ and X(0,0,1) are chosen.  
} 
\label{band1}
\end{figure}

The band structure and the density of states (DOS) for the COO-I state are shown in Fig.~\ref{band1}. In our calculations, the wave vectors throughout the first Brillouin zone of the fcc lattice were sampled. Note that orbital orderings other than the ROO-I state spontaneously break the cubic symmetry, and thus some symmetry points in the Brillouin zone of the fcc lattice become inequivalent. For example, the L point at ($\frac{1}{2},\frac{1}{2},\frac{1}{2}$) and that at (-$\frac{1}{2},\frac{1}{2},\frac{1}{2}$) are not equivalent in the COO-I state whose trigonal axis is along [111]$_{\rm c}$ shown in Fig.~\ref{coo}(a).  Within the mean-field approximation, only a particular orbital can be occupied on each sites to reduce Coulomb repulsion energy, in the large $U$ limit.   The four lowest band in the figure, indeed,  correspond to the states mainly consisting from the orbitals listed in Table~\ref{tbcoo}. The low symmetries in the occupied orbitals and the large components of hopping matrices which hybridise different kinds of orbitals, e.g., $t_{ij}^{yz,zx}$ make these
bands repulsive in each other. As a result, these bands are rather flat, disentangled and only crossing on limited symmetry points or lines.
We can find a pseudo-gap $\sim 0.2$~eV in the vicinity the Fermi level, which is only closed around the L ($\frac{1}{2},\frac{1}{2},\frac{1}{2})$ point on the trigonal axis of the COO-I state.
 
%

\begin{figure}[tb]
\begin{center}
\includegraphics[width=8.4cm]{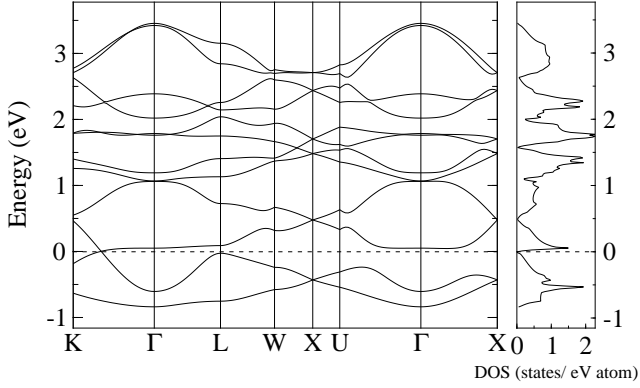}
\end{center}
\caption{
The same as Fig.~\ref{band1} but for the COO-II state obtained with $\bm{\alpha}_{\rm S}=\frac{-1}{\sqrt{2}}(1,1,0)$. 
The symmetry points K$(\frac{3}{4},\frac{3}{4},0)$, L$(\frac{1}{2},\frac{1}{2},\frac{1}{2})$, U$(\frac{1}{4},1,\frac{1}{4})$ W$(\frac{1}{2},1,0)$ and X(0,1,0) are shown.
} 
\label{band2}
\end{figure}

Figure~\ref{band2} shows the band structure and the density of states for the COO-II state. The band structure of the COO-II state is similar to that of the COO-I state (see Fig.~\ref{band1}), and band gaps at the L points which are on the direction from the centre of the Fe$_4$O$_4$ cube to the electron-rich sites are small; for instance, gaps at the L$(\frac{1}{2},\frac{1}{2},\frac{1}{2})$ and L$(-\frac{1}{2},-\frac{1}{2},\frac{1}{2})$ points are small for the COO-II state depicted in Fig.~\ref{coo}(b).
Again, a pseudo-gap in the vicinity of the Fermi level can be found among the low-lying four bands. The gap is closed only on the $\Gamma$--K symmetry line along the [110]$_{\rm c}$ direction, which is on the mirror plane of the COO-II state, and thus the COO-II state is a zero gap semiconductor. The density of states near the Fermi level reduces with increasing $V$.
Since these COO states are on the verge of gap opening, it is expected that further symmetry lowering caused by lattice distortion in the low-temperature phase would result in the formation of the insulating gap. In \S~\ref{LowT}, we will discuss the electronic state in the low-temperature phase in relation to the lattice deformations.

\section{Formation Mechanism of COO State\label{Mechanism}}%
\subsection{Interacting Fe$_4$ tetrahedral molecules of $B$ sites\label{Tetra}}%
In the previous section, we have found the COO states stabilised in large $t_{pd}$ regime in the spinless three-band  Hubbard model on the pyrochlore lattice. However, it is not obvious why such unconventional orbital orderings realise in this system. 
To gain more insight about these peculiar orbital orderings, let us consider an isolated tetrahedron of $B$ sites in infinite $U$ limit as the first step. The pyrochlore lattice can then be treated as a fcc lattice of these tetrahedral molecules of $B$ sites.

Within HF approximation, only one kind of orbital can be occupied in each of the $B$ sites, if $U=\infty$.
We further assume the solution possesses the trigonal symmetry with the [111]$_{\rm c}$ axis and is invariant with combined operator of time reversal and vertical mirror. The symmetry is the same as the COO-I state and here, we use the same site labels in Fig.~\ref{coo}(a). With these assumptions, the wave function at site 1 can take one of the $a_{1}$ or $e^{\pm}$ symmetry orbitals:
\begin{align}
|1,a_{1}\rangle &=\frac{1}{\sqrt{3}}(|1,yz\rangle +|1,zx\rangle + |1,xy\rangle) \nonumber\\
|1,e^{+}\rangle &=\frac{1}{\sqrt{3}}(\omega^2|1,yz\rangle +\omega|1,zx\rangle + |1,xy\rangle) \nonumber\\
|1,e^{-}\rangle &=\frac{-1}{\sqrt{3}}(\omega|1,yz\rangle +\omega^2|1,zx\rangle + |1,xy\rangle) \nonumber
\end{align}
and the wave function at site 2, 3 and 4 can be written as
\begin{align}
|2\rangle &=v|2,yz\rangle +u|2,zx\rangle + u^*|2,xy\rangle \nonumber  \\
|3\rangle &=u^*|3,yz\rangle +v|3,zx\rangle + u|3,xy\rangle \nonumber \\
|4\rangle &=u|4,yz\rangle +u^*|4,zx\rangle + v|4,xy\rangle, \nonumber
\end{align}
where $u$ and $v$ are complex and real constants, respectively, and satisfy the relation $2|u|^2+v^2=1$. The $a_1$ and $e^{\pm}$ symmetry molecular orbitals (MO) on the triangular cluster of sites 2, 3 and 4 are described as
\begin{align}
|234,a_{1}\rangle &=\frac{1}{\sqrt{3}}( |2\rangle + |3\rangle + |4\rangle )  \nonumber\\
|234,e^{+}\rangle &=\frac{1}{\sqrt{3}}( \omega^2 |2\rangle + \omega |3\rangle + |4\rangle )  \nonumber\\
|234,e^{-}\rangle &=\frac{-1}{\sqrt{3}}( \omega |2\rangle + \omega^2 |3\rangle + |4\rangle ) \nonumber
\end{align}
and if we express one-body energies for these $a_{1}$, $e^+$ and $e^-$ symmetry MOs as $\varepsilon^0$, $\varepsilon^{+1}$ and $\varepsilon^{-1}$, respectively, $\varepsilon^n$ can be described as 
\begin{align}
\varepsilon^n&=2\textrm{Re}[\omega^n \langle 3 | H | 2 \rangle]  \nonumber\\
  &=2\textrm{Re}[\omega^n\{- (u^*)^2 t_{\sigma} + u^2 t_{pd} + v^2 t_{pd}\}], \nonumber
\end{align}
where $H$ is the Hamiltonian described in eq.~(\ref{eq1}) and $t_\sigma=3|t_{dd\sigma}|/4$.
Here, for simplicity, the values of parameters $t_{dd\pi}$, $D$, $\zeta$ and $V$  are assumed to be zero.
Substituting $u=e^{\textrm{i}\phi}\cos\theta /\sqrt{2}$ and $v=\sin\theta$, we obtain
\begin{align}
\varepsilon^n=&- t_\sigma  \cos^2\theta\cos\left(2\phi - \frac{2\pi}{3}n\right) \nonumber \\
& + t_{pd} \cos^2\theta\cos\left(2\phi +\frac{2\pi}{3}n \right) \nonumber \\
 &+2t_{pd}\sin^2\theta\cos\left(\frac{2\pi}{3}n\right).
\end{align}
Either $e^+$ or $e^-$ symmetry MO can have the lowest energy among the three MOs on the triangular cluster:
\[
 \varepsilon^{\pm 1}=-\sqrt{t_\sigma^2+t_\sigma t_{pd} +t_{pd}^2}
 \] 
 at $\theta=0$ and 
 \[ 
  2\phi = \pm \left[\tan^{-1} \frac{\sqrt{3} t_\sigma}{2t_{pd} +t_\sigma} + \frac{\pi}{3}\right].
 \]
 The value of $2\phi$ for this lowest energy MO is within $\pi/3 < |2\phi| < 2\pi/3$ and thus orbitals on each sites are complex orbitals. For example, the orbitals at sites 2, 3 and 4 are $x^+$, $y^+$ and $z^+$ orbitals, respectively, when $\theta=0$ and $2\phi=\pi/2$ are chosen for the $e^+$ symmetry MO and this orbital alignment on the three sites is the same as that for the COO-I state in Table~\ref{tbcoo}. These MOs on the triangular cluster are further hybridised with the orbitals on site 1 having the same symmetry.
 The hopping integrals, i.e., $h^0=\langle  1,a_1 |H|234,a_1\rangle$ and $h^{\pm 1}=\langle 1,e^\pm |H|234,e^\pm\rangle$, can be written as
\begin{equation}
h^n = -t_\sigma\sin\theta \nonumber - \sqrt{2}t_{pd}\cos\theta\cos\left(\phi -\frac{2\pi}{3}n\right) \nonumber 
\end{equation}
and thus the energies of the bonding $\varepsilon^n_\textrm{b}$ and antibonding $\varepsilon^n_\textrm{a}$ states of the MO and orbital on site 1 in each symmetries can be obtained as
\begin{equation}
   \varepsilon^n_\textrm{b,a}=\frac{1}{2}\left[\varepsilon^n \pm \sqrt{ (\varepsilon^n)^2 + 4(h^n)^2 }\right].
\end{equation}

For the tetrahedron with two electrons, the lowest-energy electronic configurations are those with an electron on the $e^\pm$ symmetry MO of the triangular portion of the tetrahedron ($E^\pm$) and the other electron on the bonding MO with $e^\mp$ symmetry ($E^\mp_\textrm{b}$)  , i.e., $E^+E^-_\textrm{b}$ or  $E^-E^+_\textrm{b}$ configurations. Note that configurations such as $E^+_\textrm{b} E^-_\textrm{b}$ cost large Coulomb repulsion energy at site 1 and are not favourable.
The minimum total energy of $E^+E^-_\textrm{b}$ configuration with respect to $\theta$ and $\phi$ is found at $\theta \sim -60^\circ$ and $2\phi \sim 89^\circ$ 
with parameter values $t_{dd\sigma}=-0.275$~eV and $t_{pd}=0.4$~eV, where the COO-I state is the ground state on the pyrochlore lattice.

Now let us consider electron hopping between tetrahedral molecules on the pyrochlore lattice in addition to the internal hopping in each of them. 
 The Hamiltonian for the Bloch function with wave vector $\bm{k}$ can be easily obtained by replacing the hopping matrix of the isolated tetrahedron:
\[
   t_{ij}^{\mu\nu} \to  t_{ij}^{\mu\nu}\left[1+e^{\textrm{i}\bm{k}\cdot(\bm{r}_{j'}-\bm{r}_i)}\right],
\]
where site $j'$  is equivalent site to $j$ on the adjacent tetrahedron sharing a $\langle 110\rangle_{\rm c}$ chain with site $j$ and $\bm{r}_{i}$ ($\bm{r}_{j'}$) represents the position of the site $i$ ($j'$).
For the Bloch states with $\bm{k}$ around the $\Gamma$ point, the intra- and inter-molecular hopping terms are in-phase ($\sim 2t_{ij}^{\mu\nu}$), and thus the energy gain from the hopping term is approximately two time as large as that of the isolated tetrahedron. This stabilises a COO state, where the lowest two MOs in each tetrahedron are mainly occupied, and they are coherently aligned. On the other hand, for the Bloch states with $\bm{k}$ in the vicinity of the boundary of the first Brillouin zone, the phases of the intra- and inter-molecular hopping terms are not matched. This causes mixture of the lowest two MOs and the higher ones through the inter-molecular hopping, because of breaking of the local symmetry. This would modify the ordered MOs of the COO state from those in the isolated tetrahedron.
 
 \begin{table}
\caption{The largest four eigenvalues $w_i$ and the symmetries $\gamma_i$ and parameter values ($\theta_i$, $\phi_i$ and $\chi_i$) of corresponding eigenfunctions of the density matrix of a tetrahedral molecule in the COO-I state on the pyrochlore lattice within the HF approximation. The energies of Bloch states $\varepsilon_i$ with $\gamma_i$ symmetries at the $\Gamma$ point are also shown.  In (a), $U$=40.0~eV and $t_{dd\pi}=D=0$~eV, and  in (b), $U$=4.0~eV and $D=0.25$~eV are adopted; $t_{dp}$=0.4~eV, $V=0$~eV and the other parameters are the same as \S~\ref{HighT}.
\label{MO}}\vspace{2mm}
\begin{center}
\begin{tabular}{crrrrr}\hline
~~~~~$\gamma_i$   & $w_i$~~ &  $\theta_i$~($^\circ$)  & $2\phi_i$~($^\circ$)   &  $\chi_i$~($^\circ$)  &  $\varepsilon_i$~(eV) \\[1mm] \hline
\multicolumn {6}{l}{(a)} \\[-3mm]
~~~~~~$E^+$                      & 0.806 &  -11.7 & 93.3 &  0.3  & -1.032 \\
~~~~~~$E^-_\textrm{b}$  & 0.768 &  -12.9 & 95.0 &   -58.6  & -0.721   \\
~~~~~~$A_1$                      & 0.382 &  -11.5 & 95.0 &   0.7  &    0.008  \\
~~~~~~$E^-_\textrm{a}$  & 0.044 &  -13.1 & 92.6 &   31.3  & 1.760 \\[1mm]
\hline
\multicolumn{6}{l}{(b)} \\[-3mm]
~~~~~~$E^+$                      & 0.800 &  -10.2 & 91.3 &   4.2  & -0.791  \\
~~~~~~$E^-_\textrm{b}$  & 0.727 &  -23.2 & 112.8 &   -54.9  & -0.637   \\
~~~~~~$A_1$                      & 0.414 &  -11.2 & 89.4 &   4.1  &   -0.007  \\
~~~~~~$E^-_\textrm{a}$  & 0.036 &  -19.9 & 91.9 &   34.7  & 1.061 \\
\hline
\end{tabular}
\end{center}
\end{table}
To verify the above scenario with interacting tetrahedral molecules of $B$ sites, the density matrix $\rho_{i\mu,j\nu}=\langle c^\dagger_{j\nu}c_{i\mu}\rangle$  for a tetrahedral molecule of $B$ sites (in a $12\times 12$ size) in the COO-I state on the pyrochlore lattice is calculated and diagonalised.  Results obtained within the HF approximation with two kinds of parameter sets are shown in Table~\ref{MO}: in (a), $U$=40.0~eV and $t_{dd\pi}=D=0$, and  in (b), $U$=4.0~eV and $D=0.25$~eV; in both calculations, $t_{pd}=0.4$~eV and $V=0$~eV are adopted and the other parameters are the same as the previous section.
 The largest four eigenvalues $w_i$ and the symmetries $\gamma_i$ and parameters ($\theta_i$, $\phi_i$ and $\chi_i$) of corresponding eigenfunctions are listed.  Here, $\chi_i$ is the mixing angle between the wave functions at site 1, and the MO on the triangular part is defined as
 \[
       |\gamma_i\rangle = \sin\chi_i |1,\gamma_i \rangle +\cos\chi_i |234,\gamma_i \rangle.
  \]
 The energies of Bloch states $\varepsilon_i$ with symmetry $\gamma_i$ at the $\Gamma$ point are also shown. 
 
The values of $\theta$ and $\phi$ are almost the same among the eigenfunctions in Table~\ref{MO}(a). This is because only one particular orbital can be occupied in each sites in large $U$ limit. Variations in these parameters are larger for the results with realistic parameters  in Table~\ref{MO}(b).
However, no qualitative differences are found between the two. 
As expected, the lowest-energy two MOs $E^+$ and $E^- _\textrm{b}$ are dominantly occupied.
In addition, about 0.4 electron resides on the $A_1$ symmetry MO with the third lowest energy. This is because the lowest two MOs can be hybridised with the higher ones through the inter-molecular hopping.
The value of $\theta \sim -12^\circ$ is much smaller than that in isolated tetrahedron: $\theta \sim -60^\circ$.
This large discrepancy in the value of $\theta$ can be explained by the differences in the MO occupations.
Indeed, if the lowest-energy two MOs and the third-lowest MO are assumed to accommodate 0.8 and 0.4 electron, respectively, in the isolated tetrahedron, the configuration $(E^+)^{0.8}(E^-_\textrm{b})^{0.8}(A_1)^{0.4}$  has the minimum total energy at $\theta \sim -11^\circ$ and $2\phi \sim 94^\circ$ with parameter values $t_{dd\sigma}=-0.275$~eV and $t_{pd}=0.4$~eV.

We have described the COO-I state as the ferro-MO ordered state of the tetrahedral molecules of the $B$ sites, assuming the trigonal symmetry. Since there are similarity among the COO states, one may consider all these states can be described by a more general scheme. In the next subsection, we will find that all these COO states can be regarded as a ferromagnetic state of fictitious orbital moments of the pseudo-$d$ symmetry MOs on the tetrahedral molecules.  In this context, the differences  among them are only magnetisation directions of the ferromagnetic state.
\subsection{Stability of COO state and pseudo-$SO(3)$ symmetry\label{Stability}}
A COO state in the $e_g$ orbitals has been proposed to explain the properties of ferromagnetic metallic phase of doped manganites, where the static Jahn-Teller deformation disappears~\cite{ATakahashi,JvdBrink,Maezono}. The $e_g$ orbitals $|x^2-y^2\rangle$ and $|3z^2-r^2\rangle$ are combined with the {\it complex number} coefficients, e.g., $\frac{1}{\sqrt{2}}[|3z^2-r^2\rangle \pm \textrm{i}|x^2-y^2\rangle]$ in this COO state, and because of the isotropic charge distribution, the COO state is not associated with the Jahn-Teller deformation.  Kubo and Hirashima, however, have argued that the COO state in ferromagnetic metallic manganites is unstable, when the electron correlation effect is taken into account~\cite{KKubo}. 
We, therefore, performed calculations on a finite size system using the exact diagonalisation method to clarify whether the COO states in our three-band model are stabile under the electron correlation effect and why these peculiar COO states are preferred.

The three-band model described in \S~\ref{Model} is employed for the finite size system, where sixteen $B$ sites in the conventional fcc unit cell accommodating eight electrons are considered with the periodic condition. To reduce the number of the many-body basis functions, we assume $U=\infty$. The other parameters are the same as \S~\ref{HighT}. The Hamiltonian of the 16-site model is diagonalised with the Lanczos algorithm~\cite{EDagoto} to obtain the ground-state wave function.

\begin{figure}[tb]
\begin{center}
\includegraphics[width=8cm]{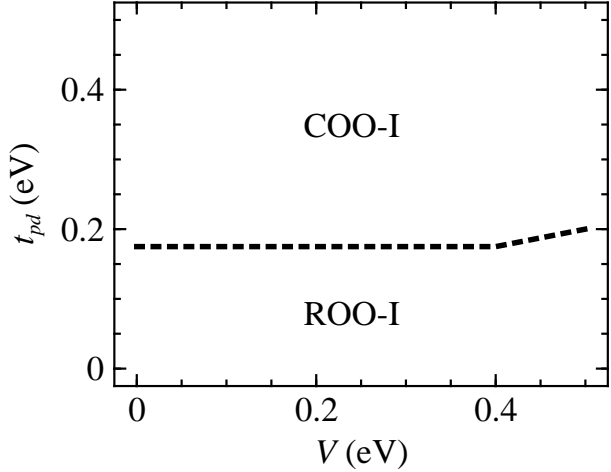}
\end{center}
\caption{
Orbital orders found in the ground state of the 16-site model obtained with the exact diagonalisation method are mapped on the $V$-$t_{pd}$ plane. In the calculations, $D=0.25$~eV and $\bm{\alpha}_{\rm S}=\frac{-1}{\sqrt{3}}(1,1,1)$ are assumed.
} 
\label{PD2}
\end{figure}

The orbital orders found in the ground state of the 16-site model are mapped on the $V$-$t_{pd}$ plane in Fig.~\ref{PD2}.  In the calculation, $D=0.25$~eV and $\bm{\alpha}_{\rm S}=\frac{-1}{\sqrt{3}}(1,1,1)$ are adopted. The regions of the ground states corresponding to the ROO-I and COO-I states can be seen. To examine instability toward the Verwey charge ordering,  the staggered potential $\sum_{i\mu} \delta_i\,n_{i\mu}$ was also applied in the  ROO-I region, where $\delta_i=+\delta$ and $\delta_i=-\delta$ for the [110]$_{\rm c}$ and [1\=10]$_{\rm c}$ chains, respectively, and $2\delta$=0.05~eV is adopted. The difference between the occupation numbers on the  [110]$_{\rm c}$ and [1\=10]$_{\rm c}$ chains continuously varies from $\sim 0.1$ to $\sim 0.3$ with increasing values of $V$ and $t_{pd}$. 
These results are consistent with the phase diagrams in Fig.~\ref{PD1} obtained from the HF calculations.

As was discussed in \S~\ref{HighT}, the energies of the three COO states are nearly degenerated within $\sim$0.01~eV, and each of them can be stabilised when $\bm{\alpha}_{\rm S}$ is oriented to the direction of their total magnetic moments. In fact, almost the same diagrams as in Fig.~\ref{PD2}  were found except the COO-I region is replaced by those for the COO-II or COO-III states with $\bm{\alpha}_{\rm S}=\frac{-1}{\sqrt{2}}(1,1,0)$ or $\bm{\alpha}_{\rm S}=(0,0,-1)$, respectively. These COO states can even continuously transform into the others as we gradually vary $\bm{\alpha}_{\rm S}$.  As discussed later, the COO states are results of spontaneous breaking of the inherent pseudo-spherical symmetry analogous to the ferromagnetic order in the Heisenberg model. 

\begin{figure}[tb]
\begin{center}
\includegraphics[width=8cm]{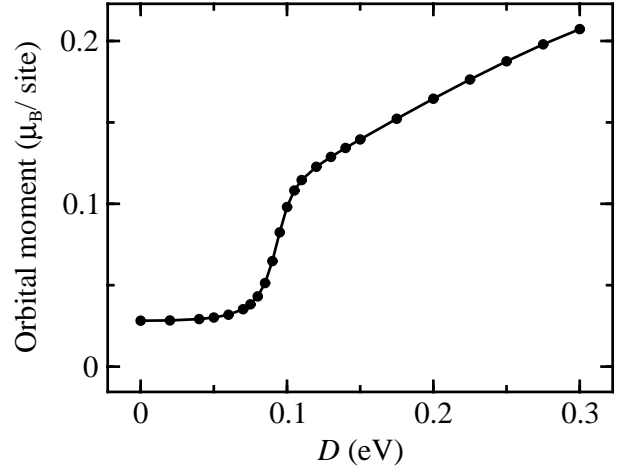}
\end{center}
\caption{
Trigonal field $D$ dependence of the total orbital moment ($\mu_\textrm{B}$ per $B$ site) obtained with $V=0.1$~eV, $t_{pd}=0.35$~eV and $\bm{\alpha}_{\rm S}=\frac{-1}{\sqrt{3}}(1,1,1)$. 
} 
\label{Moment}
\end{figure}

Figure~\ref{Moment} shows the relationship between the trigonal field $D$ and the total orbital moment per $B$ site. While large orbital moments are induced and the COO-I state is found to be the ground state for $D\gtrsim 0.1$ eV, the COO state collapses and only small orbital moments appear for small values of $D$. 
From the analysis of the $3\times 3$ density matrix $\langle c^\dagger_{i\nu}c_{i\mu}\rangle$ on each site, we found orbital occupation in each site for  $D\lesssim 0.1$~eV is described as the mixed state, where two kinds of the $e^\pi_g$ orbitals are occupied with the same probability, except for a small deviation ($\sim 10$\%) caused by the spin-orbit interaction. Thus, this ground state can be considered as an orbital-liquid state with small orbital polarisation owing to the spin-orbit interaction. On the other hand, for $D\gtrsim 0.1$~eV, one particular orbital shown in Table~\ref{tbcoo} is dominantly occupied in each site, and the occupation probabilities for the other two are small (less than 5\% for $D=0.25$~eV).
Contrastingly, the COO-I state remains to be the lowest energy solution with the HF approximation down to $D$=0. It is also noteworthy that even when the spin-orbit coupling $\zeta$ on all site is set to zero except for a site and $\bm{\alpha}_{\rm S}$ is oriented to the trigonal axis of this site, the COO-I state is still obtained with $D\gtrsim 0.1$~eV. This also shows robustness of the COO-I state for $D\gtrsim 0.1$~eV.

\begin{figure}[tb]
\begin{center}
\includegraphics[width=8cm]{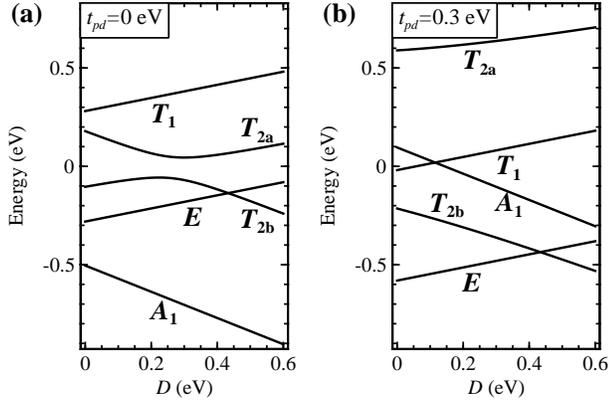}
\end{center}
\caption{Molecular-orbital levels of the isolated tetrahedral molecule of the $B$ sites as a function of $D$ obtained with $t_{pd}$=0~eV (a) and $t_{pd}$=0.3~eV (b).} 
\label{Tetra}
\end{figure}
To understand this change in the ground state as a function of $D$, let us consider an isolated tetrahedral molecule of the $B$ sites accommodating one electron. The $t_{2g}$ orbitals on four $B$ sites in the tetrahedron are reduced to five different basis sets of irreducible representations in the $T_d$ symmetry: $A_1+E+T_1+2T_2$. The MOs are classified according to the symbols of the irreducible representations, and the bases and energies are described in the Appendix. The energy levels corresponding to the bonding and antibonding states of two $T_2$ basis sets ($T^\sigma_2$ and $T^\pi_2$) are labelled $T_{2b}$ and $T_{2a}$, respectively.
In Fig.~\ref{Tetra}, MO level diagrams for the isolated tetrahedral molecule obtained with $t_{pd}=0$~eV (a) and $t_{pd}=0.3$~eV (b) are shown.

For small value of $t_{pd}$, the lowest is the $A_1$ level, where the $a_{1g}$ orbitals are occupied by 0.25 electron in each. The second lowest is the $E$ level for small values of $D$ and is the $T_{2b}$ level for $D \gtrsim 0.45$~eV. For large value of $t_{pd}$, the occupation of the $e^\pi_g$ orbitals is more favourable, and the $E$ level, where  the $e^\pi_g$ orbitals are occupied for all $B$ sites, is the lowest level. However, the $T_{2b}$ level becomes the lowest for $D\gtrsim 0.45$~eV.  

From these level diagrams, we can explain the change in the orbital ordering qualitatively. For the isolated tetrahedron with two electrons, the $E^2$ closed-shell configuration is favourable in the case that the energy separation between the lowest $E$ and higher levels is large. This costs large on-site Coulomb energy since two kind of the $e^\pi_g$ orbitals, i.e., $e_u$ and $e_v$, should be occupied on each $B$ sites (see Table~\ref{irrep}). In this situation, large quantum fluctuation is expected even on the pyrochlore lattice, since electrons should avoid occupying the same site with this limited orbital degree of the freedom. 
On the other hand, if the $T_{2b}$ and $E$ levels are nearly degenerated [$\varepsilon(T_{2b})-\varepsilon(E)\lesssim 0.2$~eV], electrons would prefer to occupy a particular orbital in each of the $B$ sites utilising both the $E$ and $T_{2b}$ basis sets to minimise the on-site Coulomb energy. 

From above discussions, the pseudo-degeneracy of the $E$ and $T_{2b}$ levels is the requisite for the formation of the COO state.  If we regard these MOs as irreducible representations of a $d$ state in the $T_d$ symmetry, i.e., $d=E+T_{2b}$,  the fictitious $d$ symmetry MOs with magnetic quantum number $m$ can be defined as
\begin{align}
 |\tilde{d},m\! =\!\pm 2\rangle &=\frac{1}{\sqrt{2}}\left( | E,v \rangle \pm \textrm{i}|T_{2b},\zeta \rangle \right), \nonumber\\
 |\tilde{d},m\! =\!\pm 1\rangle &=\frac{1}{\sqrt{2}}\left( \mp | T_{2b},\eta\rangle - \textrm{i}|T_{2b},\xi \rangle \right), \nonumber\\
 |\tilde{d},m\! =\! 0\rangle &=| E,u \rangle.  \label{pseudoD}
\end{align}
If the pseudo-$SO(3)$ symmetry in the Coulomb interactions among these $\tilde{d}$ MOs are further assumed, the most lowest energy configuration for accommodating two electrons in this spinless $\tilde{d}$ MOs is that obeys Hund's second rule: the $\tilde{d}^2$ configuration, where $|\tilde{d},m\! =\! 2\rangle$ and $|\tilde{d},m\! =\! 1\rangle$ orbitals occupied with the fictitious total orbital moments $\tilde{L}=3$.
Thus, the COO state can be interpreted as a ferromagnetic (FM) alignment of the  fictitious orbital moments of the  $\tilde{d}$ MOs, which is parallel to the real magnetisation direction of the COO states. From this viewpoint, the differences among the COO states are merely the directions of the magnetisation of the FM state.  The COO-I, COO-II and COO-III are, therefore, only three particular cases of the FM state with magnetisation directions $\langle 111\rangle_{\rm c}$, $\langle 110\rangle_{\rm c}$ and $\langle 001\rangle_{\rm c}$, respectively.

To check the validity of the assumption of the pseudo-$SO(3)$ symmetry, the density matrix $\rho_{i\mu,j\nu}$ for the tetrahedral molecule is diagonalised. For the COO-III state corresponding to the FM state with the [001]$_{\rm c}$ magnetisation direction, the largest three eigenvalues  are $w_1=0.73$, $w_2=0.54$ and $w_3=0.40$. The overlap between these eigenfunctions of $\rho$ and the $\tilde{d}$ orbitals in eq.~(\ref{pseudoD}) are $|\langle w_1|\tilde{d},m\! =\! 2\rangle|=0.96$, $|\langle w_2|\tilde{d},m\! =\! 1\rangle|=0.88$ and $|\langle w_3|\tilde{d},m\! =\! 0\rangle|=0.51$. As expected, the dominantly occupied two MOs are correspond to the $|\tilde{d},m\! =\! 2\rangle$ and $|\tilde{d},m\! =\! 1\rangle$ orbitals and their deviations from these $\tilde{d}$ orbitals are fairly small.
The COO-II state correspond to the FM state with the [110]$_{\rm c}$ direction. The wave functions of the $\tilde{d}$ states can be easily obtained by rotating the quantisation axis to the magnetisation direction using the representation matrix $D_{m,m'}[R(\alpha,\beta,\gamma)]$ of the rotation $R(\alpha,\beta,\gamma)$ described by the Eulerian angles.  For instance, the $|\tilde{d},m'\rangle$ orbital with the [110]$_{\rm c}$ quantisation axis can be obtained as
\begin{align}
 &|\tilde{d},m'\! =\! 2\rangle =\sum_{m=-2}^2 |\tilde{d},m\rangle~D_{m,2}\left[R(\pi/4,\pi/2,0)\right] \nonumber \\ 
 &~~~=\frac{\sqrt{6}}{4}| E,u \rangle -\frac{\textrm{i}}{2}\left(| T_{2b},\xi\rangle - |T_{2b},\eta \rangle \right) +\frac{1}{2\sqrt{2}}|T_{2b},\zeta \rangle, \nonumber \\
 &|\tilde{d},m'\! =\! 1\rangle = \frac{\textrm{i}}{\sqrt{2}}| E,v \rangle +\frac{1}{2}\left(| T_{2b},\xi\rangle + |T_{2b},\eta \rangle \right), \nonumber \\
 &|\tilde{d},m'\! =\! 0\rangle = -\frac{1}{2}| E,u \rangle+\frac{\sqrt{3}}{2}| T_{2b},\zeta\rangle. \nonumber 
 \end{align} 
 The largest three eigenvalues of $\rho$ in the COO-II state are $w_1=0.78$, $w_2=0.66$ and $w_3=0.39$ and the overlaps are $|\langle w_1|\tilde{d},m'\! =\! 2\rangle|=0.99$, $|\langle w_2|\tilde{d},m'\! =\! 1\rangle|=0.87$ and $|\langle w_3|\tilde{d},m'\! =\! 0\rangle|=0.39$. Again, dominantly occupied two MOs are reproduced well with the pseudo-$SO(3)$ symmetry assumption. The examination of $\rho$ in the COO-I state leads to the same conclusion, and we now realise that the $E^+$ and $E_{\rm b}^-$ MOs discussed in the previous subsection correspond to the $|\tilde{d},m\! =\! 2\rangle$ and $|\tilde{d},m\! =\! 1\rangle$ MOs, respectively.
 
 The pseudo-spherical symmetry, which permits the creation of Hund's rule state on each Fe$_4$ unit, is the fundamental ingredient for the formation and stabilisation of the COO states in the present system. In contrast,  such inherent symmetry has not been discussed and seems to be absent in the $e_g$ system for the ferromagnetic manganites~\cite{ATakahashi,JvdBrink,Maezono,KKubo}. 
\section{COO State in the Low-Temperature Phase\label{LowT}}
%
\begin{figure}[t]
\begin{center}
\includegraphics[width=7cm]{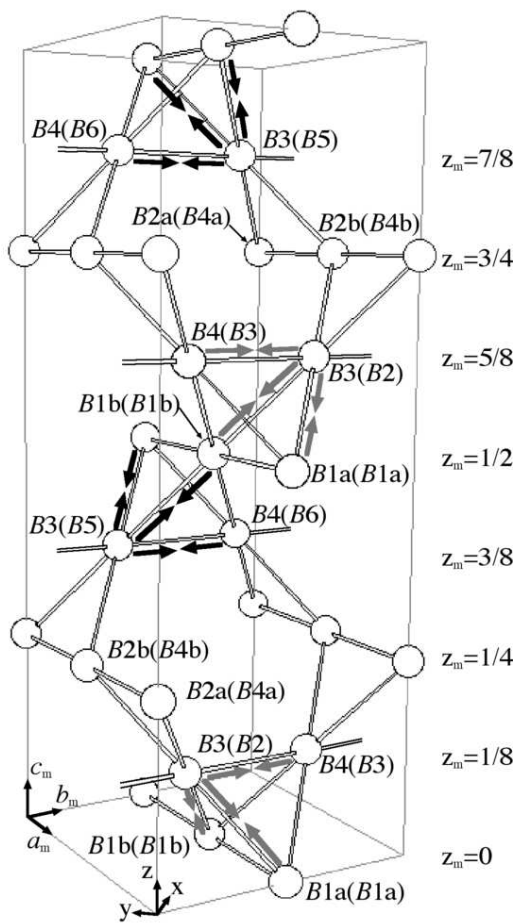}
\end{center}
\caption{
Illustration of an $a_{\rm c}/\sqrt{2}\times a_{\rm c}/\sqrt{2}\times 2a_{\rm c}$ cell for the {\it Pmca} and {\it Pmc}\,2$_1$ pseudo-orthorhombic structures~\cite{WrightPRB,MIizumi}. The circles represent Fe ions on the $B$ sites. The six inequivalent $B$ sites in the {\it Pmca} structural model are labelled $B1$a, $B1$b, $B2$a, $B2$b, $B3$ and $B4$ after ref.~\citen{WrightPRB}, and those in the parentheses  are for the {\it Pmc}\,2$_1$ structural model in ref.~\citen{MIizumi}. 
Prominently contracted Fe-Fe bonds $\sim 0.1$~{\AA} in the {\it Pmca} (those with the black and grey arrows) and in the {\it Pmc}\,2$_1$ (those with the black arrows) structural models are also shown. The monoclinic $a_{\rm m}$-, $b_{\rm m}$- and $c_{\rm m}$-axes are along the [\=1\=10]$_{\rm c}$, [1\=10]$_{\rm c}$ and [001]$_{\rm c}$ directions, respectively.
} 
\label{Distortions}
\end{figure}
\subsection{Lattice distortion and related properties in the low-temperature phase}
In the low-temperature phase ($T<T_{\rm V}$), the lattice distortion to the monoclinic ({\it Cc}) or triclinic ({\it P}1) symmetries with a cell size $\sqrt{2}a_{\rm c}\times \sqrt{2}a_{\rm c}\times 2a_{\rm c}$ occurs. However, attempts to determine the true crystal structure have been hindered by overlapping of the Bragg peaks owing to the presence of pseudo-symmetries.
As an approximated crystal-structure model for the low-temperature phase, Iizumi \textit{et al.} proposed an $a_{\rm c}/\sqrt{2}\times a_{\rm c}/\sqrt{2}\times 2a_{\rm c}$ subcell imposing orthorhombic symmetry constraints on the atomic positions to reduce the number of free parameters from the true $\sqrt{2}a_{\rm c}\times \sqrt{2}a_{\rm c}\times 2a_{\rm c}$ {\it Cc} supercell accommodating 16 inequivalent {\it B} sites in their neutron diffraction study~\cite{MIizumi}. They used the centric  {\it Pmca} and acentric {\it Pmc}\,2$_1$ space groups for candidates of orthorhombic pseudo-symmetry. Recently, Wright \textit{et al.} refined the crystal structure of the low-temperature phase from x-ray and neutron diffraction experiments~\cite{WrightLett,WrightPRB}. The same crystal structure model with the orthorhombic {\it Pmca} pseudo-symmetry is applied for the analysis. The inequivalent sites are reduced from 16 sites to 6 sites within the approximation, and they are labelled $B1$a, $B1$b, $B2$a, $B2$b, $B3$ and $B4$ (see Fig.~\ref{Distortions}). Jahn-Teller-like distortions of the oxygen octahedra surrounding the $B1$ sites are found below $T_{\rm V}$ from their experiments, and charge disproportionation is inferred from variations in the Fe-O bond lengths.

The electronic state in the low-temperature phase has been investigated using LDA + {\it U} calculations~\cite{ILeonovLett, HTJengLett, ILeonovFull, HTJengFull} based on the monoclinic lattice structure shown by Wright \textit{et al}~\cite{WrightPRB}. From the LDA + {\it U} calculations, a state with simultaneous charge and antiferro-orbital ordering is obtained. In the state, the $B2$a, $B2$b and $B3$ sites are vacant, and the $xy$ orbitals are occupied at the $B4$ sites. The $yz$ and $zx$ orbitals align alternately along the $B1$ chains, where the Jahn-Teller-like distortions are observed. 

The cell distortion of magnetite in the low-temperature phase is closely approximated by that with a rhombohedral symmetry~\cite{MIizumimono,LJVieland,WrightRhombo}, although the atomic displacements in the cell have pseudo-orthorhombic symmetry~\cite{WrightPRB}.
The monoclinic distortion in the low-temperature phase has been conventionally interpreted as the result of the combination of the microscopic orthorhombic distortion caused by charge ordering and macroscopic rhombohedral distortion owing to the magnetostriction.
The magnetic easy axes are oriented along the [111]$_{\rm c}$ direction in the high-temperature phase and the monoclinic $c_{\rm m}$-axis (more precisely, it tilts toward the [111]$_{\rm c}$ direction by $2^{\circ}$~\cite{KAbe}) in the low-temperature phase~\cite{ARMuxworthy}. Here the monoclinic $a_{\rm m}$-, $b_{\rm m}$- and $c_{\rm m}$-axes are along the [\=1\=10]$_{\rm c}$, [1\=10]$_{\rm c}$ and [001]$_{\rm c}$ directions, respectively. Interestingly, the easy axis switching accompanied by simultaneous change of $c_{\rm m}$-axis is observed by applying external magnetic field below $T_{\rm V}$~\cite{switching}, and the dependence of the Verwey transition temperature under the application of magnetic field was also reported~\cite{MZiese,DCMertens}.  Moreover, the magnetoelectric effects which indicate triclinic distortion in the low-temperature phase were also discussed~\cite{GTRado1,GTRado2,KSiratori,YMiyamotoME1,YMiyamotoME2,KKato}.

As was discussed in the previous section, the COO state is a kind of magnetic order within the orbital degree of freedom. This means that its orbital polarisation depends on the magnetisation directions and thus strong coupling between the magnetism and lattice distortions, i.e., magnetostriction, is anticipated. Because of this strong coupling with the lattice, the COO state in the high-temperature phase would be modified by the monoclinic lattice distortion below $T_{\rm V}$.  In the remainder of this section, we will discuss the relationship between the COO state and lattice distortion. A COO state expected to be stabilised by the monoclinic lattice distortion in the low-temperature phase will be also presented.

\subsection{Relationship between lattice distortions and COO states\label{hopanddis}}
%
\begin{table}[t]
\caption{
Expressions in terms of $t_\sigma$ and $t_{pd}$ and the first derivative with respect to $t_\sigma$ of the absolute value of the hopping integrals between the dominantly occupied orbitals $h_{ij}=\langle i,\mu|H_t|j,\nu\rangle$ in the COO-I, COO-II and COO-III states. $|i,\mu\rangle$ denotes the dominantly occupied orbital $\mu$  at site $i$ in the notation in Fig.~\ref{coo}. 
Here, $t_{\sigma}=3|t_{dd\sigma}|/4$, and $t_{dd\pi}$ is neglected. $H_t$ is the hopping term in eq.~(\ref{eq1}).
}
\vspace{2mm}
\begin{center}
\extrarowheight=0.3mm
\begin{tabular}{cccc}\hline
state & $h_{ij}$  & $|h_{ij}|$  & $\frac{d\,|h_{ij}|}{d\,t_{\sigma}}$ \\ 
\hline
\raisebox{-3.5mm}[0pt][0pt]{COO-I} & $\langle 3,y^+|H_t|2,x^+\rangle$  & $\frac{1}{2}(t_{\sigma}+t_{pd})$&0.50\\ 
& $\langle 4,z^+|H_t|1,e^-\rangle$ & $ \sqrt{\frac{2+\sqrt{3}}{6}}t_{pd}$&0\\
\hline
& $\langle 3,110|H_t|2,110\rangle$ & $\frac{1}{3}(2t_{\sigma}+t_{pd})$ & 0.67\\
\raisebox{-1mm}[0pt][0pt]{COO-II} & $\langle 3,110|H_t|4,z^+\rangle$  & $\frac{1}{\sqrt{12}}(t_\sigma+2t_{pd})$ & 0.29\\
& $\langle 4,z^+|H_t|1,z^-\rangle$ & $ t_{pd}$ & 0\\[1mm]
\hline
\raisebox{-2.5mm}[0pt][0pt]{COO-III} & $\langle 2,\textrm{1\=11}|H_t|1,\textrm{\=1\=11}\rangle$ & $\frac{1}{3}(t_{\sigma}+2t_{pd})$ & 0.33\\ 
&  $\langle 4,\textrm{111}|H_t|1,\textrm{\=1\=11}\rangle$ & $\frac{1}{3}(t_{\sigma}+t_{pd})$ & 0.33 \\
\hline
\end{tabular}
\end{center}
\label{hop} 
\end{table}
%

To discuss the relationship between the lattice deformation and COO states, the hopping integrals between the dominantly occupied orbitals $h_{ij}=\langle i,\mu|H_t|j,\nu\rangle$ in the COO-I, COO-II and COO-III states are examined.
Here, $|i,\mu\rangle$ represents the dominantly occupied orbital $\mu$ at site $i$ which is listed in Table~\ref{tbcoo}. If the absolute value of the hopping integral is increased by the Fe--Fe bond contraction, the energy of corresponding bonding state is decreased, and this would reinforce the stability of the ordering. In Table~\ref{hop}, $|h_{ij}|$ expressed  in terms of $t_\sigma$ and $t_{pd}$, and the first derivative of $|h_{ij}|$ with respect to $t_\sigma$ are listed. Here, $t_\sigma=3|t_{dd\sigma}|/4$ is the absolute value of the hopping integral of the $\sigma$-bond, which is most sensitive to the change in the bond length among the hopping integrals. Note that there are the relations $h_{32}$=$h_{24}$=$h_{43}$ and $h_{14}$=$h_{12}$=$h_{13}$ for the COO-I state, $h_{34}$=$h_{42}$=$h_{21}$=$h_{13}$ for the COO-II and COO-III states, and $h_{14}$=$h_{23}$ for the COO-III state.

Judging from the derivatives, the Fe--Fe bond between the two electron-poor sites of the COO-II state, on which orbitals $|\textrm{110}\rangle$ are occupied, is most sensitive to the variation of $t_{\sigma}$. Thus, the orthorhombic distortion, in which the Fe--Fe bond along the [1\=10] direction is contracted, is expected to stabilise the COO-II state.
Similarly, the COO-I state is accompanied by the rhombohedral lattice distortion; three Fe--Fe bonds on the triangle of sites 2, 3 and 4 of each of the Fe$_4$O$_4$ cubes are compressed,  and this leads to elongation of the cubic cell along the [111]$_{\rm c}$ diagonal direction. On the other hand, only small tetrahedral distortion is expected for the COO-III state.

\subsection{Hartree-Fock calculations for the low-temperature phase}
The spinless three-band Hubbard model described in eq.~(\ref{eq1}) was investigated to find the electronic state realised in magnetite below $T_{\rm V}$. The calculations were carried out within the HF approximation [eq.~(\ref{HF})]. In order to introduce the effects of the lattice deformation for the low-temperature phase, we considered the variation in the values of hopping integrals as a function of Fe--Fe bond length $r$. For simplicity, we take into account such $r$ dependence on the value of the hopping integral for the $\sigma$-bond $-t_\sigma=3t_{dd\sigma}/4$, which is  most sensitive to the change in $r$ among the hopping integrals. The value of hopping integral is approximately proportional to $r^{-5}$, and we expand $t_\sigma$ up to the first order of the deviation $\delta r$ from the Fe--Fe distance in the high-temperature phase $r_{\rm c}=2.966$~{\AA} as
\[ 
t_\sigma = t^c_\sigma(1-5\beta_\sigma \delta r/r_{\rm c}),
\]
where $t^c_\sigma=0.206$~eV is the hopping integral for the high-temperature phase.
A constant parameter $\beta_\sigma$ is introduced to change the extent of the distortion effects for the $\sigma$-bonds, and we assumed $\beta_{\sigma}=3$ in the calculations.

The $a_{\rm c}/\sqrt{2}\times a_{\rm c}/\sqrt{2}\times 2a_{\rm c}$ orthorhombic unit cell consists of 16 $B$ sites (see Fig.~\ref{Distortions}) is assumed. For simplicity, effects of atomic displacements in the cell are considered only as variations in hopping integrals, and the monoclinic tilt of the $c_{\rm m}$-axis toward the $a_{\rm m}$-axis are neglected. Note that the pairs of sites labelled with `a' and `b', e.g., $B1$a and $B1$b, are equivalent without this monoclinic tilt of the $c_{\rm m}$-axis because of recovery of the $b_{\rm m}$-$c_{\rm m}$ mirror plane. Two sets of the atomic coordinates with the centric {\it Pmca}~\cite{WrightPRB} and acentric {\it Pmc}\,2$_1$~\cite{MIizumi} symmetries were used to evaluate the values of $t_\sigma$, and those adopted in the HF calculations are listed in Table~\ref{tbhopdis}. The direction of the minority spin moment is assumed to be along the $c_{\rm m}$-axis, i.e., $\bm{\alpha}_{\rm S}=(0,0,-1)$ since the magnetic easy axis for $T<T_{\rm V}$ is approximately along this direction. The other parameters are the same as those used in the calculations for the high-temperature phase: $D=0.25$~eV$, U=4.0$~eV and $\zeta=0.052$~eV. The wave vectors throughout the first Brillouin zone of the orthorhombic lattice were sampled in the HF calculations.

\begin{table}[t]
\caption{
Deviation of the bond length from that in the high-temperature phase $\delta r$ and corresponding value of the hopping integral $t_{\sigma}~(=3|t_{dd\sigma}|/4)$ adopted in the HF calculations are shown for each of the Fe--Fe bonds among the $B$ sites. Those evaluated from two sets of atomic coordinates in the {\it Pmca} \cite{WrightPRB} and {\it Pmc}\,2$_1$ \cite{MIizumi} structural models with $\beta_{\sigma}=3$ are presented.
}
\vspace{2mm}
\begin{center}
\begin{tabular}{lrllrl}\hline
\multicolumn{3}{c}{{\it Pmca}} &\multicolumn{3}{c}{{\it Pmc}\,2$_1$} \\
~bond& $\delta r$ (\AA) & $t_{\sigma}$~(eV) & ~bond& $\delta r$ (\AA) &$t_{\sigma}$~(eV) \\ \hline
$B1$--$B1$ &  0.000 &  0.206 & $B1$--$B1$ & -0.046 &  0.255 \\
$B1$--$B3$ & -0.107 &  0.318 & $B1$--$B1$ &  0.046 &  0.158 \\
$B1$--$B4$ & -0.013 &  0.220 & $B1$--$B2$ & -0.050 &  0.259 \\
$B2$--$B2$ &  0.000 &  0.206 & $B1$--$B3$ &  0.052 &  0.152 \\	 
$B2$--$B3$ &  0.080 &  0.123 & $B1$--$B5$ & -0.118 &  0.330 \\
$B2$--$B4$ &  0.042 &  0.163 & $B1$--$B6$ & -0.008 &  0.215 \\
$B3$--$B4$ & -0.082 &  0.292 & $B2$--$B3$ & -0.002 &  0.209 \\	
$B3$--$B4$ &  0.083 &  0.120 & $B2$--$B3$ &  0.005 &  0.201 \\
              &  &                     &$B2$--$B4$ & -0.004 &  0.211 \\	
              &  &                     &$B3$--$B4$ & -0.065 &  0.274 \\
              &  &                     &$B4$--$B4$ & -0.079 &  0.289 \\
              &  &                     &$B4$--$B4$ & 0.079 &  0.123 \\		
              &  &                     &$B4$--$B5$ & 0.117 &  0.085 \\	
              &  &                     &$B4$--$B6$ & 0.080 &  0.123 \\	
              &  &                     &$B5$--$B6$ & -0.104 &  0.315 \\
              &  &                     &$B5$--$B6$ &  0.105 &  0.097 \\		
\hline
\end{tabular}
\end{center}
\label{tbhopdis} 
\end{table}
%
\begin{figure*}[ht]
\begin{center}
\includegraphics[width=16cm]{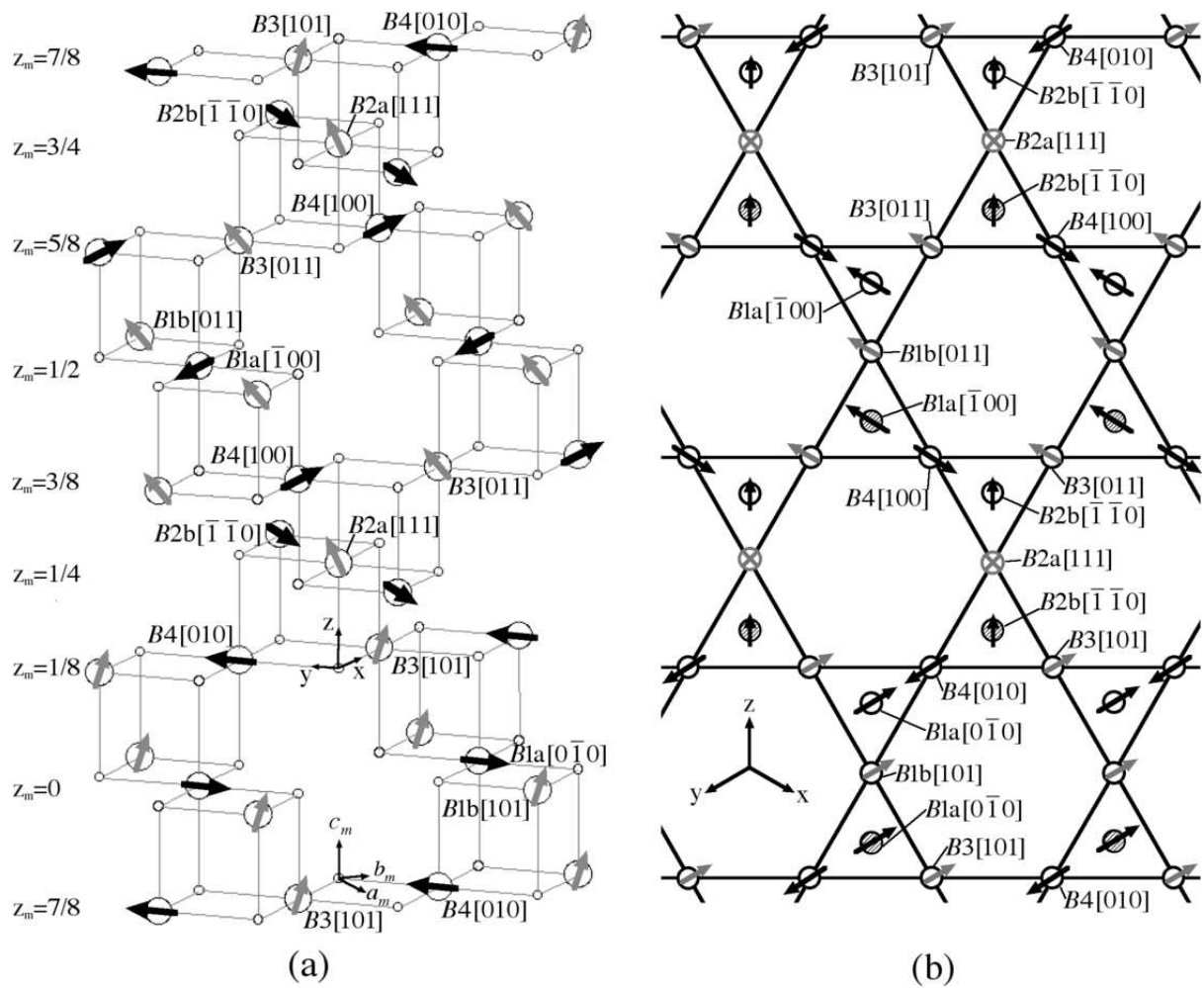}
\end{center}
\caption{
(a) Schematic representation of the COO-IV state found in the HF calculations for the low-temperature phase. The large and small circles denote the Fe ions on the $B$ sites and oxygen ions, respectively. The black and grey arrows signify the directions of orbital moments  with large and small values of occupation number at the {\it B} sites, respectively.  (b) The COO-IV state projected on the [111]$_{\rm c}$ plane. The circles represent the Fe ions on the $B$ sites. The open and hatched circles inside the triangles are above and below kagom\'e plane, respectively, and the others are on the kagom\'e lattice. The site labels in the {\it Pmca} symmetry are also shown.
} 
\label{MCOOII}
\end{figure*}
\begin{table*}[t]
\caption{
Details of the COO-IV state obtained from the calculations with the hopping integrals evaluated from the {\it Pmca} and {\it Pmc}\,2$_1$ structural models.  The $z$-coordinate of the atomic position ($z_{\rm m}$), dominantly occupied orbital and  its orbital moment (in units of $\mu_{\rm B}$), and occupation numbers (for $V=0.3$~eV and $t_{pd}=0.35$~eV) on each of the 16 sites are listed.
}
\vspace{2mm}
\begin{center}
\extrarowheight=0.3mm
\begin{tabular}{clcll}\hline
$z_{\rm m}$ & orbital [ $\omega=\exp({2\pi \textrm{i}/3})$ ]& moment & $\langle n\rangle$ ({\it Pmca}) & $\langle n\rangle$ ({\it Pmc}\,2$_1$)   \\ \hline
0 & $|\textrm{101}\rangle=\frac{1}{\sqrt{6}}[2|zx\rangle-\textrm{i}|xy\rangle+\textrm{i}|yz\rangle]$& ~~$\frac{2}{3}(1,0,1)$  & 0.411 ($B1$b) &0.418 ($B1$b) \\
0 & $|y^{-}\rangle~=\frac{1}{\sqrt{2}}[|xy\rangle+\textrm{i}|yz\rangle]$ & ~~~~~$(0,-1,0)$& 0.617 ($B1$a) &0.612 ($B1a$)   \\  
$\frac{1}{8}$ & $|\textrm{101}\rangle=\frac{1}{\sqrt{6}}[2|zx\rangle-\textrm{i}|xy\rangle+\textrm{i}|yz\rangle]$ & ~~$\frac{2}{3}(1,0,1)$& 0.434 ($B3$) &0.463 ($B2$)  \\ 
$\frac{1}{8}$ & $|y^{+}\rangle~=\frac{1}{\sqrt{2}}[|xy\rangle-\textrm{i}|yz\rangle]$ & ~~~~~$(0,1,0)$ & 0.539 ($B4$) &0.519 ($B3$)   \\   
$\frac{1}{4}$ & $|\textrm{111}\rangle=\frac{1}{\sqrt{3}}[|xy\rangle+\omega^2|yz\rangle+\omega|zx\rangle]$& ~~$\frac{1}{\sqrt{3}}(1,1,1)$ & 0.486 ($B2$a) &0.453 ($B4$a)   \\ 
$\frac{1}{4}$ & $|\textrm{\=1\=10}\rangle=\frac{1}{\sqrt{6}}[2|xy\rangle+\textrm{i}|yz\rangle-\textrm{i}|zx\rangle]$& $-\frac{2}{3}(1,1,0)$ & 0.540 ($B2$b) &0.570 ($B4$b)  \\ 
$\frac{3}{8}$ & $|\textrm{011}\rangle=\frac{1}{\sqrt{6}}[2|yz\rangle+\textrm{i}|xy\rangle-\textrm{i}|zx\rangle]$ & ~~$\frac{2}{3}(0,1,1)$ & 0.434 ($B3$) &  0.432 ($B5$) \\ 
$\frac{3}{8}$ & $|x^{+}\rangle~=\frac{1}{\sqrt{2}}[|zx\rangle-\textrm{i}|xy\rangle]$ & ~~~~~$(1,0,0)$& 0.539 ($B4$) & 0.533 ($B6$)   \\ 
$\frac{1}{2}$ & $|x^{-}\rangle~=\frac{1}{\sqrt{2}}[|zx\rangle+\textrm{i}|xy\rangle]$ & ~~~~~$(-1,0,0)$& 0.617 ($B1$a) & 0.612 ($B1$a)   \\ 
$\frac{1}{2}$ & $|\textrm{011}\rangle=\frac{1}{\sqrt{6}}[2|yz\rangle+\textrm{i}|xy\rangle-\textrm{i}|zx\rangle]$ & ~~$\frac{2}{3}(0,1,1)$ & 0.411 ($B1$b) & 0.418 ($B1$b)  \\ 
$\frac{5}{8}$ & $|\textrm{011}\rangle=\frac{1}{\sqrt{6}}[2|yz\rangle+\textrm{i}|xy\rangle-\textrm{i}|zx\rangle]$ & ~~$\frac{2}{3}(0,1,1)$ & 0.434 ($B3$) &0.463 ($B2$)  \\
$\frac{5}{8}$ & $|x^{+}\rangle~=\frac{1}{\sqrt{2}}[|zx\rangle-\textrm{i}|xy\rangle]$ & ~~~~~$(1,0,0)$& 0.539 ($B4$) & 0.519 ($B3$)   \\ 
$\frac{3}{4}$ & $|\textrm{111}\rangle=\frac{1}{\sqrt{3}}[|xy\rangle+\omega^2|yz\rangle+\omega|zx\rangle]$ &  ~~$\frac{1}{\sqrt{3}}(1,1,1)$& 0.486 ($B2$a) &0.453 ($B4$a)   \\
$\frac{3}{4}$ & $|\textrm{\=1\=10}\rangle=\frac{1}{\sqrt{6}}[2|xy\rangle+\textrm{i}|yz\rangle-\textrm{i}|zx\rangle]$ & $-\frac{2}{3}(1,1,0)$& 0.540 ($B2$b) &0.570 ($B4$b)  \\ 
$\frac{7}{8}$ & $|y^{+}\rangle~=\frac{1}{\sqrt{2}}[|xy\rangle-\textrm{i}|yz\rangle]$ & ~~~~~$(0,1,0)$ & 0.539 ($B4$) & 0.533 ($B6$)  \\   
$\frac{7}{8}$ & $|\textrm{101}\rangle=\frac{1}{\sqrt{6}}[2|zx\rangle-\textrm{i}|xy\rangle+\textrm{i}|yz\rangle]$& ~~$\frac{2}{3}(1,0,1)$ & 0.434 ($B3$) &0.432 ($B5$)  \\ 
\hline
\end{tabular}
\end{center}
\label{DetailMCOOIIP2_c} 
\end{table*}

From the both calculations using the hopping integrals evaluated from the {\it Pmca} and {\it Pmc}\,2$_1$ structural models, we obtained a charge and complex-orbital ordered state (COO-IV). Schematic representation of the COO-IV state is shown in Fig.~\ref{MCOOII}, and its details are listed in Table~\ref{DetailMCOOIIP2_c}. For the both symmetries, the COO-IV state is found as the lowest energy solution in almost the same area on $V$-$t_{pd}$ plane where the COO phases presence in Fig.~\ref{PD1}(c). Although we assumed the orthorhombic lattice structure, the symmetry of the COO-IV state is spontaneously lowered from the {\it Pmca} and {\it Pmc}\,2$_1$ to the monoclinic $P\underline{2}/\underline{c}$ and $P\underline{c}$, respectively. However, the symmetry of the COO-IV state is essentially $P\underline{2}/\underline{c}$ and deviation from this is small in the calculation with the {\it Pmc}\,2$_1$ lattice symmetry. Thereby, for the remainder of this paper, we refer the $B$ sites by the labels for the {\it Pmca} lattice symmetry.

As seen in Fig.~\ref{MCOOII}, the COO-IV state can be regarded as an alternative stacking of microscopic layers (thickness $a_{\rm c}$) of the COO-II states with two different orientations: those having easy-axes along the [101]$_{\rm c}$ and  [011]$_{\rm c}$ directions.  The horizontal planes at the $z$-coordinate of the lattice $z_{\rm m}=1/4$ and  $z_{\rm m}=3/4$ correspond to `domain walls', on which  the $B2$ sites are located. Each of the $B2$a sites is placed at the intersection of a [01\=1]$_{\rm c}$ chain of the $|011\rangle$ orbitals and a [10\=1]$_{\rm c}$ chain of the $|101\rangle$ orbitals, and its orbital moment is along the  [111]$_{\rm c}$ direction.
 Similarly, each of the $B2$b sites is positioned at a intersection of a [011]$_{\rm c}$ chain of the $|x^\pm \rangle$ orbital and a [101]$_{\rm c}$ chain of the $|y^\pm\rangle$ orbitals, and its orbital moment is oriented along the [\=1\=10]$_{\rm c}$ direction.
The $B$1-$B$3 bonds are the most contracted bonds, and both the chains of the $|011\rangle$ and $|101\rangle$ orbitals of the COO-II layers are on these bonds. From the discussions in \S~\ref{hopanddis}, this explains why the COO-IV state is stabilised with the lattice distortion in the low-temperature phase.

The magnetic easy axis of the COO-IV is oriented to the [223]$_{\rm c}$ direction tilted from the $c_{\rm m}$-axis toward the $[111]_{\rm c}$ direction about 43$^\circ$ and the total orbital moment $\sim 0.2~\mu_\textrm{B}$ per $B$ site. The experimental easy axis is also tilted from the $c_{\rm c}$-axis toward the same direction below $T_{\rm V}$. However, the tilting angle is about 2$^\circ$.~\cite{KAbe}

The charge disproportionation in the COO-IV state is rather moderate similar to the COO-II state, and the charge order satisfies the Anderson condition, which constrains every tetrahedron to have two electrons.  From the bond valence sums analysis of the x-ray and neutron diffraction experiments~\cite{WrightLett,WrightPRB}, charge ordering with apparent local charges of Fe$^{2.4+}$ and Fe$^{2.6+}$ has been found in the low-temperature phase. Although the small magnitude of charge splitting is consistent with our results, the charge order found in the analysis does not satisfy the Anderson condition.

\begin{figure}[ht]
\begin{center}
\includegraphics[width=8.4cm]{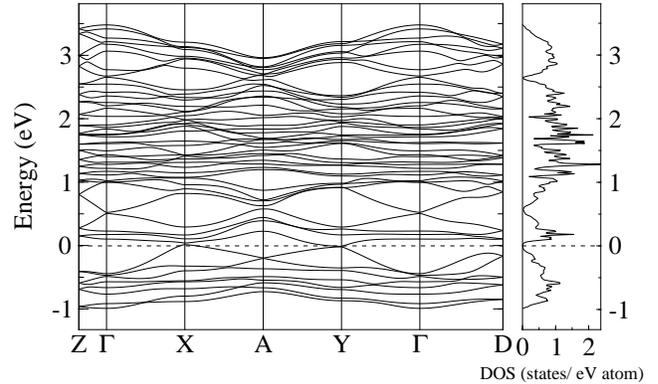}
\end{center}
\caption{
Band structure and DOS of the COO-IV state obtained from calculations with $V=0.3$~eV, $t_{pd}=0.35$~eV and the hopping integrals evaluated from the {\it Pmca} atomic positions. 
The symmetry points Z$(0,0,\frac{1}{2})$, X$(\frac{1}{2},0,0)$, A$(\frac{1}{2},\frac{1}{2},0)$, Y$(0,\frac{1}{2},0)$ and  D$(\frac{1}{2},0,-\frac{1}{2})$ with units $\frac{2\pi}{a_{\rm m}}=\frac{2\pi}{b_{\rm m}}=\frac{2\sqrt{2}\pi}{a_{\rm c}}$ and $\frac{2\pi}{c_{\rm m}}=\frac{\pi}{a_{\rm c}}$ are chosen. The dotted line represent the Fermi level. The band gap at the Fermi level does not open.
} 
\label{BDPmca}
\end{figure}

The band structure and DOS of the COO-IV state for the {\it Pmca} symmetry are shown in Fig.~\ref{BDPmca}. For the {\it Pmca} symmetry, the band gap at the Fermi energy does not open; it is closed at the Y$(0,\frac{1}{2},0)$ point and mostly closed at the X$(\frac{1}{2},0,0)$ point. The band gaps at the L points which are on the direction from the centre of the Fe$_4$O$_4$ cube to the electron-rich sites are small in the COO-II state, and corresponding bands at the L$(\frac{1}{2},\frac{1}{2},\frac{1}{2})$ point of the two COO-II micro-layers come to the X$(\frac{1}{2},0,0)$ point and those of the L$(-\frac{1}{2},\frac{1}{2},\frac{1}{2})$ and L$(\frac{1}{2},-\frac{1}{2},\frac{1}{2})$ points are positioned at the Y$(0,\frac{1}{2},0)$ point. In contrast to the results of the calculations with the {\it Pmca} lattice symmetry,  the COO-IV state obtained assuming the {\it Pmc}\,2$_1$ symmetry has the band gap at the Fermi level, and its magnitude is $\sim2.5\times10^{-2}$~eV for $V=0.3$~eV and $t_{pd}=0.35$~eV. However, the band gap does not open for the small values of $V$.
 
%

A $\sqrt{2}a_{\rm c}\times \sqrt{2}a_{\rm c}\times 2a_{\rm c}$ supercell comprising 64 {\it B} sites, which is the same dimension to the {\it Cc} symmetry cell,  is also examined using the hopping integrals obtained from the {\it Pmc}\,2$_1$ atomic coordinates.  However, the COO-IV state is obtained as the ground state, and no further COO state is found. In this supercell, the X and Y points of the {\it Pmca} or {\it Pmc}\,2$_1$ cells correspond to the $\Gamma$ point, and the energy gap at the Fermi level is most narrow at this symmetry point in the first Brillouin zone. 

\section{Discussions}\label{Discussions}
The electronic state in the high-temperature phase is still in controversy. Anderson proposed the state in which two Fe$^{3+}$ and two Fe$^{2+}$ ions are equally contained in every {\it B} site tetrahedron with the short-range ordering.~\cite{Anderson}  Yamada \textit{et al.} have suggested a molecular polaron model~\cite{YYamada,YYamada2}. On the other hand, half-metallic state has been predicted in LSDA calculations~\cite{ZZhang,AYanaseAPW,AYanaseFLAPW}, and results of resonant x-ray scattering measurements support the itinerant-electron picture\cite{NoCOhigh}. 

From the results of the HF calculation with the cubic structure, the COO-I state is anticipated to realise in the high-temperature phase. This is consistent with the fact that the magnetic easy axis for the high-temperature phase is in the $\langle 111\rangle_{\rm c}$ direction. However, the rhombohedral distortion expected in the presence of the COO-I state is not observed. Yamada {\it et al.} proposed a molecular polaron model to explain the neutron diffuse scattering found in the high-temperature phase~\cite{YYamada,YYamada2}. In this model, the valence fluctuation of Fe$_4$O$_4$ cubic cluster is strongly coupled with dynamical lattice  distortion, i.e., the $T_{2g}$ mode, within the cluster.
 However, the local lattice distortions assumed in the model (see Fig.~1 in ref.~\citen{YYamada}) are the same as those in the presence of the COO-I and COO-II states as we discussed in \S~\ref{hopanddis}. Thus the diffuse scattering can be interpreted as a result of the fluctuation of the fictitious orbital moment on the Fe$_4$O$_4$ clusters strongly coupled with dynamical lattice distortion. For this reason, we consider the COO-I state remains short-range order in the high-temperature phase and thus no statical rhombohedral distortion occurs in the high-temperature phase.

As for the low-temperature phase, variety of charge ordering model have been proposed~\cite{VerweyCO,MMizoguchi,JMZuo}. However, the NMR~\cite{NoCOlow} and x-ray anomalous scattering~\cite{NoCOlow2} cast doubts on the existence of charge ordering. Later on, moderate charge orderings with apparent local charges of Fe$^{2.4+}$ and Fe$^{2.6+}$ are inferred from the x-ray and neutron diffraction experiments~\cite{WrightLett,WrightPRB}. The COO-IV state obtained for the low-temperature phase is also accompanied by charge disproportionation of Fe$^{2.6+}$ and Fe$^{2.4+}$.  However, the pattern of charge ordering predicted from the present theory is not the same to those inferred from x-ray and neutron diffraction experiments. 

\begin{figure}[t]
\begin{center}
\includegraphics[width=8.5cm]{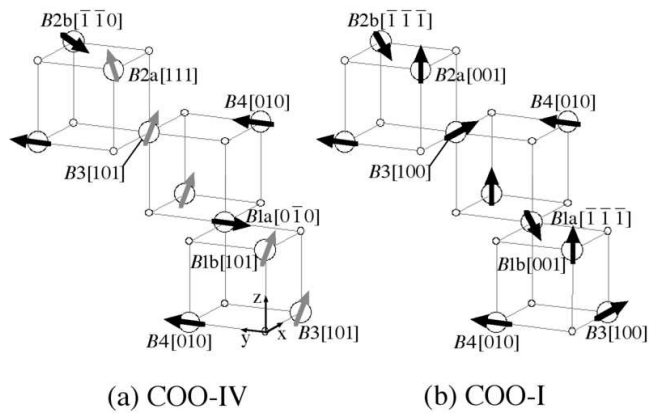}
\end{center}
\caption{
Continuous transformation of the COO-IV state (a) into the COO-I state (b) with increasing value of $t_{pd}$ from  $t_{pd}=0.5$~eV. Alignments of orbital moments within $-1/8 \le z_{\rm m} \le 1/4$ are shown.
} 
\label{COOIVtoCOOI}
\end{figure}

We have described the COO-IV state as an alternative stacking of microscopic layers of the COO-II state with two different orientations. However, the state can also be regarded as a COO-I state with the [111]$_{\rm c}$ magnetisation direction modulated by the lattice distortion: the orbital magnetisation direction is slanted from direction [111]$_{\rm c}$ to [011]$_{\rm c}$ in the layer $1/4 < z_{\rm m} <3/4$ and direction [111]$_{\rm c}$ to [101]$_{\rm c}$ in the layer $-1/4 < z_{\rm m} <1/4$. Indeed, the COO-IV state can even continuously transform into the COO-I state as the values of $t_{pd}$ is increased from $t_{pd}=0.5$~eV with $V=0$~eV as demonstrated in Fig.~\ref{COOIVtoCOOI}.  As was discussed in \S~\ref{hopanddis},  the COO-I state is expected to be accompanied by the rhombohedral cell distortion. This explains why the distortion of the unit cell in the low-temperature phase can be approximated by the rhombohedral distortion. 

\begin{figure}[t]
\begin{center}
\includegraphics[width=7cm]{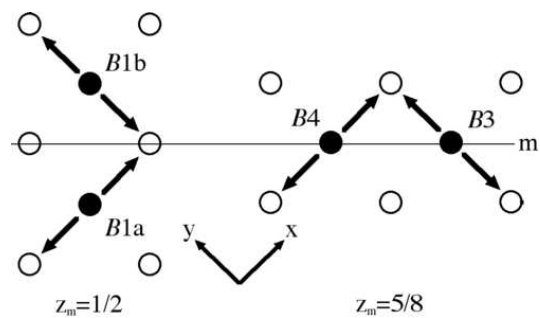}
\end{center}
\caption{
Illustration of the Jahn-Teller-like distortions along the $B1$ chain at $z_{\rm m}=1/2$ (on the left) and the $B3$/$B4$ chain at $z_{\rm m}=5/8$ (on the right).
The pseudo-mirror plane perpendicular to the $a_{\rm m}$-axis presumed in the crystal structure analysis is also shown.
The open and closed circles represent oxygen and Fe ions, respectively. The arrows from Fe ions represent the elongated Fe-O bonds. 
The Jahn-Teller-like distortions along the $B1$ chain are observed in the experiments~\cite{WrightPRB}, and those along the $B3$/$B4$ chain, which break the mirror symmetry, are also anticipated in the presence of the COO-IV state.
} 
\label{Jahn}
\end{figure}

As was mentioned before, the Jahn-Teller like distortions of oxygen octahedra surrounding the Fe ions at the $B1$ sites are observed in the experiments~\cite{WrightPRB}.
This distortion found in the low-temperature phase can be account for the presence of the COO-IV ordering as follows.
The Fe ions at the $B1$a and $B1$b sites and the neighbouring oxygens on the $z_{\rm m}=1/2$ plane are illustrated on the left side of Fig.~\ref{Jahn}.
The Fe--O bonds along the $x$-axis are elongated, and those along the $y$-axis are contracted at the $B1$a site. The Fe--O lengths along the two directions are reversed at the $B1$b site. The complex orbitals at the $B1$a and $B1$b sites are $|x^-\rangle$ and $|011\rangle$ in the COO-IV state (see Table~\ref{DetailMCOOIIP2_c}) and their electron clouds are stretched along the $x$- and $y$-axes, respectively. This antiferro-orbital alignment along the $B1$ chain is, therefore, stabilised by the Jahn-Teller like distortion with alternative directions of the elongated  Fe--O bonds along the chain. The same holds for the $B1$ chains on the $z_{\rm m}=0$ plane with the $|101\rangle$/$|y^-\rangle$ antiferro-orbital alignment.

The similar antiferro-orbital alignment of $|011\rangle$/$|x^+\rangle$ or  $|y^+\rangle$/$|101\rangle$ are present along the $B3$/$B4$ alternative chains parallel to the $b_{\rm m}$-axis in the COO-IV state. The alternative Jahn-Teller-like distortion is anticipated along the $B3$/$B4$ chain from the theory. However, such distortions are still not found in the x-ray or neutron diffraction experiments. This chains are on the pseudo-mirror plane perpendicular to the $a_{\rm m}$-axis presumed in the crystal structure analysis (see Fig.~\ref{Jahn}) and search for such distortion on the $B3$/$B4$ chains can be a check for the validity of the present theory.

From results of the high-resolution photoemission spectroscopy (PES) experiments~\cite{JHPark}, the insulating gap with the magnitudes $\sim 0.1$~eV and $\sim 0.15$~eV are inferred above and below $T_{\rm V}$, respectively.  Other PES experiments~\cite{DSchrupp} and oxygen $K$-edge absorption measurements~\cite{EGering} also support non-closure of the gap above $T_{\rm V}$ and indicate small-polaronic behaviour of charge carriers. The formation of the pseudo-gap with negligible Drude weight in $T_{\rm V}<T<300$~K and a clear opening of the optical gap of $\sim 0.14$~eV below $T_{\rm V}$ are observed in measurements of the optical-conductivity spectra~\cite{SKPark}.
  The COO states obtained with the cubic structure are either a semimetal with extremely low carrier density or a zero-gap semiconductor and have a pseudo-gap with the magnitude $\sim 0.2$~eV in the vicinity of Fermi level. In addition, the COO state is expected to be short-range order due to the strong coupling with the dynamical lattice distortion in the high-temperature phase. Thus, small-polaronic or incoherent-metallic behaviours of the high-temperature phase inferred from the above experiments are consistent with the present theory.
  
Since the energy gap at the Fermi level is almost opened in these COO states obtained with the cubic structure, symmetry lowering caused by the monoclinic lattice distortion in the low-temperature phase would result in the formation of the insulating gap. 
Indeed, the COO-IV state has a insulating gap of $\sim 0.02$~eV with the {\it Pmc}\,2$_1$ symmetry and large value of $V\sim 0.3$~eV.
Interestingly, the absence of the inversion symmetry leads to the COO-IV state being multiferroic\cite{KSiratori}.
For the {\it Cc} supercell, the band gap at the Fermi level is narrow or closed only at the $\Gamma$ point. The size of this gap might be larger, if the rhombohedral cell distortion in combination with the {\it Cc} atomic displacements are further considered.

\section{Conclusions\label{Conclusions}}
In this paper, a spinless three-band Hubbard model for $t_{2g}$ electrons with minority spin on the $B$ sites is investigated with the Hartree-Fock (HF) approximation and  exact diagonalisation method to discuss the electronic structure and Verwey transition in magnetite. 

Complex-orbital ordered (COO) states with noncollinear orbital moments $\sim 0.4~\mu_{\rm B}$ are found as the ground state with the cubic structure of the high-temperature phase when the electron hopping via adjacent O 2$p$ orbitals is strong. Two kinds of COO states with total orbital moment along directions $\langle 111 \rangle_{\rm c}$ (COO-I state) and $\langle 110 \rangle_{\rm c}$  (COO-II state) are obtained. These states are semimetal or zero-gap semiconductor, and pseudo-gaps of $\sim 0.2$~eV are formed near the Fermi level.  

The formation mechanism of the COO state is discussed analytically and the stability of the state against the electron correlation effects is examined with the exact diagonalisation method for the 16-site system. The COO state does not stem from the orbital moments induced by ordered spin moments via the spin-orbit interaction as usual magnetic materials. It is an exotic magnetic ordering within the orbital degree of freedom.  The system of electrons with minority spin on the $B$ sites can be described as a fcc lattice of interacting Fe$_4$ tetrahedral molecules. From this viewpoint, the COO state originates from the formation of Hund's second rule state in spinless pseudo-$d$ symmetry molecular orbitals on each of the Fe$_4$ tetrahedron and ferromagnetic alignment of their fictitious orbital moments.

For the low-temperature phase, the HF calculations were performed with the hopping integrals evaluated from the {\it Pmca} and {\it Pmc}\,2$_1$ structural models. A COO state, which can be  described as an alternative stacking of microscopic layers of the COO-II states with two different $\langle 110 \rangle_{\rm c}$ directions of the magnetisation, is obtained (COO-IV state). The state is accompanied by moderate charge disproportion of Fe$^{2.6+}$ and Fe$^{2.4+}$ and in accordance with Jahn-Teller like distortion. Although we used the orthorhombic atomic positions, the symmetry of the COO-IV state spontaneously lowers to the monoclinic. Since the state can also be regarded as a modulated COO-I state with pseudo-trigonal symmetry, this explains conventional interpretation of the monoclinic distortion: combination of the rhombohedral cell distortion and atomic displacements in the cell with the orthorhombic symmetry.

From these findings, we interpret the Verwey transition as follows.
Above $T_\textrm{V}$, the COO-I state remains to be short-range order impeded by  dynamical lattice distortion. The strong coupling between the COO-II states and lattice leads to the formation of the COO-IV state below $T_\textrm{V}$.
\section*{Acknowledgment}
We appreciate the Supercomputer Center, Institute for Solid State Physics, University of Tokyo, for the facilities and the use of the Hitachi SR11000. This work is partly supported by a Grant-in-Aid for Scientific Research from the Ministry of Education, Culture, Sports, Science and Technology.

\appendix
\section{MOs in Isolated Tetrahedron}
The four $t_{2g}$ orbitals of the isolated tetrahedral molecule can be reduced to five different basis sets of the irreducible representations, i.e., $A_1+E+T_1+2T_2$ in the $T_d$ point group symmetry.
The energies $\varepsilon(\Gamma)$ and bases $|\Gamma,\gamma\rangle$ of these irreducible representations $\Gamma$ are listed in Table~\ref{irrep}.
There are two different basis sets for the $T_2$ irreducible representation and here, we call them $T^\sigma_2$ and $T^\pi_2$.
The hopping integral between them $\langle T_2^\sigma | H | T_2^\pi \rangle$ is also shown. For each of the basis sets of the three dimensional representations $T_1$ and $T_2$, only one of the bases is presented; the other two bases, i.e., those with $\beta$ and $\gamma$ for $T_1$ and $\eta$ and $\zeta$ for $T_2$,  are easily obtained from the bases in the table by rotating $\pm 2\pi/3$ around the [111]$_{\rm c}$ axis of the tetrahedron.
\begin{center}
\begin{table}
\caption{
Energies $\varepsilon(\Gamma)$ and bases $|\Gamma,\gamma\rangle$ of irreducible representations $\Gamma$ in the $T_d$ symmetry on the $t_{2g}$ orbitals
of tetrahedral molecule.
}
\label{irrep}
\vspace{2mm}
\extrarowheight=0.3mm
\begin{tabular}{lrl}\hline
$\Gamma$ & \multicolumn{2}{l}{Energies $\varepsilon(\Gamma)$ and bases $|\Gamma,\gamma\rangle$} \\
\hline
$A_1$&$\varepsilon(A_1)$ &= $\frac{3}{4}t_{dd\sigma} -2t_{dd\pi}+2t_{pd}$, \\[1.5mm]
&$|A_1\rangle$ &$=\frac{1}{2}\big[\frac{1}{\sqrt{3}}(+|1,yz\rangle +|1,zx\rangle + |1,xy\rangle )$ \\
&                    &~~~~$+\frac{1}{\sqrt{3}}(+|2,yz\rangle -|2,zx\rangle - |2,xy\rangle )$ \\
&                     &~~~~$+\frac{1}{\sqrt{3}}(-|3,yz\rangle +|3,zx\rangle - |3,xy\rangle )$ \\
&                    &~~~~$+\frac{1}{\sqrt{3}}(- |4,yz\rangle -|4,zx\rangle + |4,xy\rangle )~ \big]$ \\
\hline
$E$ &$\varepsilon(E)$ &= $\frac{3}{4}t_{dd\sigma} -\frac{1}{2}t_{dd\pi} - t_{pd}$, \\[1.5mm]
&$|E,u\rangle$ &$=\frac{1}{2}\big[\frac{1}{\sqrt{6}}(-|1,yz\rangle -|1,zx\rangle + 2|1,xy\rangle )$ \\
&                    &~~~~$+\frac{1}{\sqrt{6}}(-|2,yz\rangle +|2,zx\rangle - 2|2,xy\rangle )$ \\
&                     &~~~~$+\frac{1}{\sqrt{6}}(+|3,yz\rangle -|3,zx\rangle - 2|3,xy\rangle )$ \\
&                    &~~~~$+\frac{1}{\sqrt{6}}(+ |4,yz\rangle +|4,zx\rangle + 2|4,xy\rangle )~ \big]$, \\[1.5mm] 
&$|E,v\rangle$ &$=\frac{1}{2}\big[\frac{1}{\sqrt{2}}(|1,yz\rangle -|1,zx\rangle  )$ \\
&                   &~~~~$+\frac{1}{\sqrt{2}}(|2,yz\rangle +|2,zx\rangle )$ \\
&                     &~~~~$-\frac{1}{\sqrt{2}}(|3,yz\rangle +|3,zx\rangle )$ \\
&                    &~~~~$-\frac{1}{\sqrt{2}}(|4,yz\rangle -|4,zx\rangle )~ \big]$ \\
\hline
$T_1$ &$\varepsilon(T_1)$ &= $-\frac{3}{4}t_{dd\sigma} +\frac{1}{2}t_{dd\pi} - t_{pd}$, \\[1mm]					
&					$|T_1,\alpha\rangle$ &$=\frac{1}{2}\big[\frac{1}{\sqrt{2}}(|1,zx\rangle -|1,xy\rangle  )$ \\
&                    &~~~~$-\frac{1}{\sqrt{2}}(|2,zx\rangle -|2,xy\rangle )$ \\
&                     &~~~~$-\frac{1}{\sqrt{2}}(|3,zx\rangle +|3,xy\rangle )$ \\
&                   &~~~~$+\frac{1}{\sqrt{2}}(|4,zx\rangle +|4,xy\rangle )~ \big]$ \\
\hline
$T_2$ &$\varepsilon(T_2^\sigma)$ &= $-\frac{1}{4}t_{dd\sigma} +\frac{2}{3}t_{dd\pi} - \frac{2}{3}t_{pd}$, \\[1mm]
&$\varepsilon(T_2^\pi)$ &= $~~\frac{1}{4}t_{dd\sigma} -\frac{1}{6}t_{dd\pi} +\frac{5}{3} t_{pd}$, \\[1mm]
&\multicolumn{2}{l}{$\langle T_2^\sigma | H | T_2^\pi \rangle =\frac{2\sqrt{2}}{3} (\frac{3}{4}t_{dd\sigma}+t_{dd\pi} -t_{pd})$,}	\\[2mm]
&$|T_2^\sigma,\xi\rangle$ &$=-\frac{1}{2}\big[\frac{1}{\sqrt{3}}(+|1,yz\rangle +|1,zx\rangle + |1,xy\rangle )$ \\
&                     &~~~~~~$+\frac{1}{\sqrt{3}}(+|2,yz\rangle -|2,zx\rangle - |2,xy\rangle )$ \\
&                    &~~~~~~$-\frac{1}{\sqrt{3}}(-|3,yz\rangle +|3,zx\rangle - |3,xy\rangle )$ \\
&                    &~~~~~~$-\frac{1}{\sqrt{3}}(- |4,yz\rangle -|4,zx\rangle + |4,xy\rangle )~ \big]$, \\[1.5mm]				
&$|T_2^\pi,\xi\rangle$ &$=-\frac{1}{2}\big[\frac{1}{\sqrt{6}}(2|1,yz\rangle - |1,zx\rangle - |1,xy\rangle )$ \\
&                     &~~~~~~$+\frac{1}{\sqrt{6}}(2|2,yz\rangle +|2,zx\rangle + |2,xy\rangle )$ \\
&                    &~~~~~~$+\frac{1}{\sqrt{6}}(2|3,yz\rangle +|3,zx\rangle - |3,xy\rangle )$ \\
&                    &~~~~~~$+\frac{1}{\sqrt{6}}(2|4,yz\rangle -|4,zx\rangle + |4,xy\rangle )~ \big]$ \\
\hline
\end{tabular}
\end{table}
\end{center}

%
\end{document}